\documentclass[useAMS,usenatbib]{mn2e}
\usepackage{graphicx,amsmath,txfonts}

\title[Migration of accreting planets in radiative discs]{Migration of accreting planets in radiative discs from dynamical torques}
\author[A. Pierens, S.N Raymond]{A. Pierens $^{1},$ S.N. Raymond $^{1}$ \\
$^1$Laboratoire d'astrophysique de Bordeaux, Univ. Bordeaux, CNRS, B18N, all{\'e}e Geoffroy Saint-Hilaire, 33615 Pessac, France}
\date{Released 2012 Xxxxx XX}

\pagerange{\pageref{firstpage}--\pageref{lastpage}} \pubyear{2014}

\def\LaTeX{L\kern-.36em\raise.3ex\hbox{a}\kern-.15em
    T\kern-.1667em\lower.7ex\hbox{E}\kern-.125emX}
\begin{document}
\label{firstpage}
\maketitle
\begin{abstract} 
We present the results of hydrodynamical simulations of the orbital evolution of planets undergoing 
runaway gas accretion in radiative discs. We consider accreting disc models with constant 
mass flux through the disc, and where radiative cooling balances the effect of viscous heating and stellar 
irradiation. We assume that 20-30 $M_\oplus$ giant planet cores are formed in the  region where viscous heating 
dominates and migrate outward under the action of a strong entropy-related corotation torque. In the 
case where gas accretion is neglected and for an $\alpha$ viscous stress parameter $\alpha=2\times 10^{-3}$, we find evidence for strong
 dynamical torques in accreting discs 
with accretion rates ${\dot M}\gtrsim 7\times 10^{-8} \;M_\odot/yr$. Their main effect is to  increase outward 
migration rates by a factor of $\sim 2$ typically. In the presence of gas accretion, however, runaway outward migration is observed with the planet passing through the zero-torque radius and  the transition between 
the viscous heating and stellar heating dominated regimes.  The ability for an accreting planet to enter a fast migration regime 
is found to depend strongly on the planet growth rate, but can occur for values of the mass flux through the 
disc  of ${\dot M}\gtrsim 5\times 10^{-8} \;M_\odot/yr$. We find that an episode of runaway outward migration can cause an accreting planet formed in
the  5-10 AU region to temporarily orbit at star-planet separations as large as $\sim$60-70 AU. However,  increase in the amplitude of the Lindblad torque associated  with planet growth plus  change in the streamline topology near the planet  systematically cause the direction of migration to be reversed. Subsequent evolution corresponds to the planet migrating inward rapidly until  it becomes massive enough to open a gap in the disc and migrate in the 
Type II regime. Our results indicate that a planet can reach large orbital distances under the combined 
effect of dynamical torques and gas accretion, but an alternative mechanism is required to explain the presence of massive planets 
on wide orbits. 
\end{abstract}

\section{Introduction}
Radial velocity surveys indicate that giant planets exist around $\sim 14 \%$ of Sun-like stars (Cumming et al. 2008; Mayor et al. 2011), 
with 
a frequency  that is strongly correlated to the metallicity of the host star.  There exists a great diversity of orbital 
architectures among this population of 
exoplanets, which span a range of eccentricities and periods from days to  several thousands of days. The abundance 
of  giant planets is found to depend strongly on the star-planet orbital separation. For example, hot Jupiters with semi major axes $\lesssim 0.5$ AU are found to orbit around only $0.5-1 \%$ of main sequence stars,  whereas there is an excess of Jupiter-mass planets located between 1-2 AU 
(Cumming et al. 2008).  Giant planets orbiting beyond the snow-line at  5-10 AU from their star also appear to be common, and 
microlensing surveys  (Shvartzvald et al. 2016) even suggest that half of the stars may host such planets. 
Giant planets with mass $\ge 5 M_J$ have  also been observed at larger orbital distances 10-100 AU by direct imaging (Brandt et al. 2014). The 
system HR8799 (Marois et al. 2008, 2010) provides an example of such a planetary system that has been directly imaged. It contains four giant planets with mass 
in the range $7-10M_J$ and located at orbital distances $15-70$ AU. Interestingly, it has been suggested that these planets may be engulfed 
in  8:4:2:1 Laplace resonances (Gozdziewski \& Migaszewski 2014; Maire et al. 2015) that guarantee the stability of the system. \\
 It has been proposed that the planets 
  observed by direct imaging  may have formed by gravitational instability 
(Dodson-Robinson et al. 2009) or through pebble accretion (Lambrechts et al. 2014).  Due to the difficulty for forming cores that are massive enough to undergo runaway gas accretion at distances $\gtrsim30$ AU,  explaining the in-situ 
formation of massive 
planets on wide orbits through core accretion does indeed appear challenging.  In the context of this scenario, however, it can not be excluded
 that  these planets formed at distances $\lesssim$  $20$ AU where the accretion timescales are short compared to the lifetime 
of the disc, and then migrated outward. Such a process arises for example  when two planets are in mean motion resonance and evolve within 
a common gap. Provided that the inner planet is more massive than the outer one and that the disc viscosity is small enough, Crida et al. (2009a) 
have shown that a pair of planets formed in the 5-20 AU region may reach distances as large as 100 AU through outward migration. \\
A single planet can also reach large semi major axes due to the action of dynamical corotation torques (Paardekooper 20114),  which  
operate in protoplanetary discs with low viscosity  and a few times more 
massive than the Minimum Mass Solar Nebula (hereafter MMSN; Hayashi 1981).  The dynamical corotation torque   results from  the torque exerted by the region of 
the disc 
that is bound to the planet plus the torque due to the gas material that flows across the orbit as the planet migrates. In isothermal 
discs,  the dynamical  torque scales with the vortensity gradient inside the disc and the planet drift rate (Paardekooper 2014), with the possibility of 
runaway migration in the case where the planet migrates in the direction set by the corotation torque.  A similar effect can arise in 
non-isothermal discs, where the entropy-related corotation torque is the main agent for driving outward migration. Provided the latter is strong enough 
to allow for outward migration, the region  trapped to the planet tends to be  gas depleted and the dynamical torque restricts to the orbit-crossing torque which 
has a positive feedback on migration. This can make the planet reaching high drift rates and eventually  migrating through  zero-torque radii where 
analytical formulae for the disc torque (Paardekooper et al. 2011) predict stalling of migration. For example, Pierens (2015) found that a Neptune-mass 
planet embedded in a 5 MMSN radiative disc can migrate from 5 AU up to $\sim 30$ AU by this process, while analytical estimations would predict 
stopping of migration at $\sim 10$ AU.\\
In this paper, we extend our previous study by including the effect of gas accretion onto the planet. The aims are to i) investigate how 
the effects of dynamical torques on the planet orbital evolution are impacted by this process and ii) examine whether massive planets evolving on wide orbits can be 
formed by the combined effect of dynamical torques and gas accretion.  With respect to our previous work, we also consider accreting disc models 
with constant mass flow through the disc (Bitsch et al. 2014), in order to estimate the stages of disc evolution during which dynamical torques can be important. 
Our results suggest that gas accretion has two opposite effects on dynamical torques. It tends to make runaway outward migration easier due to the 
onset of non-linear effects, but planet growth ultimately leads to a reversal of migration due to the continuous increase of the Lindblad torque and change of the 
streamline topology near the planet. This makes it difficult for massive planets at large distances from their star to be formed by this process. \\
  This paper is organized as follows. In Sect. 1, we present the numerical setup. In Sect. 2, we describe the structure of the accreting 
  disc models we employ. In Sect. 3, we present the results of simulations for non-accreting cores and discuss the 
  effects of gas accretion in Sect. 4.  Finally,  we draw our conclusions in Sect. 5.

\section{Model set-up}
\subsection{Numerical method}
\label{sec:num}

 Hydrodynamical simulations were performed using the GENESIS  numerical code which solves 
the governing equations for the disc evolution on a polar grid.  This code 
employs an advection scheme based on the monotonic  transport 
algorithm (Van Leer 1977) and includes the FARGO algorithm (Masset 2000) to avoid time step limitation due to the 
Keplerian velocity at the inner edge of the disc.  The energy equation that is implemented in the code includes the effect of 
viscous heating, stellar irradiation, and radiative cooling. It reads:
\begin{equation}
\frac{\partial e}{\partial t}+\nabla \cdot (e{\bf v})=-(\gamma-1)e{\nabla \cdot {\bf v}}+Q^+_{vis}-Q^--2H\nabla  \cdot {\bf F}
\label{eq:energy}
\end{equation}
where $e$ is the thermal energy density, $\bf v$ the velocity, $\gamma$ the adiabatic index which is set 
to $\gamma=1.4$, and $H$ is the disc scale height. In the previous equation,  $Q^+_{vis}$ is the viscous heating term,  and $Q^-=2\sigma_B T_{eff}^4$ is the local radiative cooling from the disc surfaces, where 
$\sigma_B$ is the Stephan-Boltzmann constant and $T_{eff}$ the effective temperature which is given by (Menou \& Goodman 2004):
\begin{equation}
T_{eff}^4=\frac{T^4-T_{irr}^4}{\tau_{eff}} \quad \text{with} \quad \tau_{eff}=\frac{3}{8}\tau+\frac{\sqrt{3}}{4}+\frac{1}{4\tau}
\end{equation}
Here, $T$ is the midplane temperature and $\tau=\kappa\Sigma/2$ is the vertical optical depth, where $\Sigma$ is the surface density and $\kappa$  the Rosseland mean opacity which is taken from Bell \& Lin (1994) and computed assuming a constant solid/gas ratio $Z=0.01$. $T_{irr}$ is the irradiation temperature which is computed from the irradiation flux (Menou \& Goodman 2004):
\begin{equation}
\sigma_B T_{irr}^4=A \frac{L_\star(1-\epsilon)}{4\pi R^2}\frac{H}{R}\left(\frac{d \log H}{d\log R}-1\right)
\end{equation}
 where $\epsilon=1/2$ is the disc albedo and $L_\star$  the stellar luminosity. The factor $d \log H/d\log R$ is set to 
be $d \log H/d\log R=9/7$ (Chiang \& Goldreich 1997), which implies that self-shadowing effects are not taken into account in this study. 
Provided that an embedded planet does not carve a deep gap in the disc, test simulations have shown that this leads to temperature and surface density structure of accreting disc models that are in very good agreement with 
the 3D hydrodynamical models of Bitsch et al. (2014). \\
In Eq. \ref{eq:energy}, ${\bf F}$ is the radiative flux which is treated in the flux-limited  diffusion approach and which 
reads (e.g. Kley \& Crida 2008):
\begin{equation}
{\bf F}=-\frac{16\sigma \lambda T^3}{\rho \kappa}\nabla T
\end{equation}
where $\rho=\Sigma/2H$ is the midplane density and  where $\lambda$ is a flux-limiter (e.g.  Kley 1989).\\

In this work, we do not take into account the effects of the disc self-gravity.  For the most massive discs we consider, this could artificially  alter the torque 
felt by a migrating planet.   In the case where self-gravity is not included, disc gravity indeed causes  a change in the angular velocity of the planet only, resulting in a spurious shift of Lindblad resonances (Pierens \& Hur\'e 2005). To force both the disc and the planet to 
evolve in the same gravitational potential, we therefore  subtract the axisymmetric component of the disc gravity 
prior to evaluating the disc torque exerted on the planet. Such a  procedure results in migration rates that are consistent with fully self-gravitating 
calculations (Baruteau \& Masset 2008), and has been already employed in previous simulations of migrating planets undergoing gas accretion (D'angelo \& Lubow 2008).    

When calculating the torque exerted by 
the disc on the planet,  we also exclude the material contained within a distance $0.6R_H$ from the planet, where $R_H$ is the 
planet Hill radius. Such a value is consistent with the one recommended by Crida et al. (2009b) to obtain 
an accurate estimation of the migration rate. Moreover,    
the potential of the planet is smoothed over a distance $b=0.4H_p$, where $H_p$ is the disc scale height at the location of the planet.  
Using $b=0.4H_p$ leads to linear torques that are in good agreement with those derived from a 3D linear theory (Paardekooper et al. 2010). Also, 
we note that it has been shown that for planets with planet-to-star mass ratios similar to those considered here, 2D simulations that 
employ such a value for the softening length provide a correct estimation of the half-width of the horseshoe region derived 
from 3D calculations (Fung et al. 2015), and therefore reasonably estimate the non-linear horseshoe drag for such planets.\\

Gas accretion onto the protoplanet is modelled by removing  a fraction of the gas located within half of the Hill sphere during each time step, 
and by adding this mass to the planet. The gas removal rate is chosen such that the accretion timescale onto the planet is $t_{acc}=f_{red}t_{dyn}$, where 
$t_{dyn}$ is the orbital period of the planet and $f_{red}$ is a free parameter for which we considered values of  
$f_{red}=1,5,10$. The case $f_{red}=1$ corresponds to the  maximum rate at which the planet can accrete gas material from the disc (Kley 1999).\\

The computational domain is covered by $N_R=1024$ radial grid cells uniformly distributed  between $R_{in}=2$  and 
$R_{out}=30$ AU typically, and $N_\phi=1024$ azimuthal grid cells. For simulations in which gas accretion is included, the 
disc extends up to $60$ AU and we employ a number of radial grid cells twice as large.

\subsection{Initial conditions}
\label{sec:initial}

We consider  accreting disc models with  constant mass flow through the disc $\dot{M}$, with $\dot{M}$  given by: 
\begin{equation}
\dot M=3\pi \nu \Sigma
\label{eq:mdot}
\end{equation}
where $\nu$ is the viscosity.  Here, the effective kinematic viscosity  is modeled using the standard alpha prescription $\nu=\alpha h^2R^2\Omega$, with $\alpha=2\times 10^{-3}$ and where 
$h=H/R$ is the disc aspect ratio. In the outer parts of the disc where stellar heating dominates over viscous heating, we expect 
$h\propto R^{2/7}$ (Chiang \& Goldreich 1997)  and this leads to, using Eq. \ref{eq:mdot}, $\Sigma=\Sigma_0 (R/R_0)^{-15/14}$ for a disc with  constant $\dot M$
(Bitsch et al. 2014). These profiles for the aspect ratio and surface density of the disc correspond to the ones that are adopted as initial conditions in our simulations. \\ 

The surface density 
$\Sigma_0$ at $R=R_0$ is determined according to the chosen $\dot {M}$, for which we consider values of $\dot M =3\times 10^{-8}, 5\times 10^{-8}, 
7\times 10^{-8}, 10^{-7} \; M_\odot/yr$.   For our reference disc with $\dot{M}=7\times 10^{-8}\; M_\odot/yr$, 
$\Sigma_0=1600 g/cm^2$ at $R_0=5$ AU.  For this value of the accretion rate, the stellar luminosity is set to $L_\star=1.43 L_\odot$ where $L_\odot$ is the 
solar luminosity.   However, to account for the change in luminosity due to stellar evolution, we follow Bitsch et al. (2015) and consider the stellar luminosity 
to depend on the value for $\dot{M}$  we impose. The value for $L_\star$ as a function of $\dot{M}$ is taken from Bitsch et al. (2015).

\subsection{Boundary conditions}

To obtain a  disc model with  steady accretion flow through the disc, we allow inflow of disc material at the inner boundary while we  impose the value 
for the accretion rate $\dot{M}$ at the outer edge of the computational domain (Bitsch et al. 2015). This is done by employing a wave-killing zone at the 
outer boundary (De Val-Borro et al. 2006) where both the surface density and the radial velocity are relaxed toward their initial value 
(Kley \& Haghighipour 2014). Since our initial condition for the surface density is such that $\Sigma \propto R^{-15/14}$, this implies that $R_{out}$ must be large 
enough so that the disc structure in the wave-killing region  is indeed determined by stellar irradiation.

\section{Structure of the accreting disc models}
\begin{figure}
\centering
\includegraphics[width=\columnwidth]{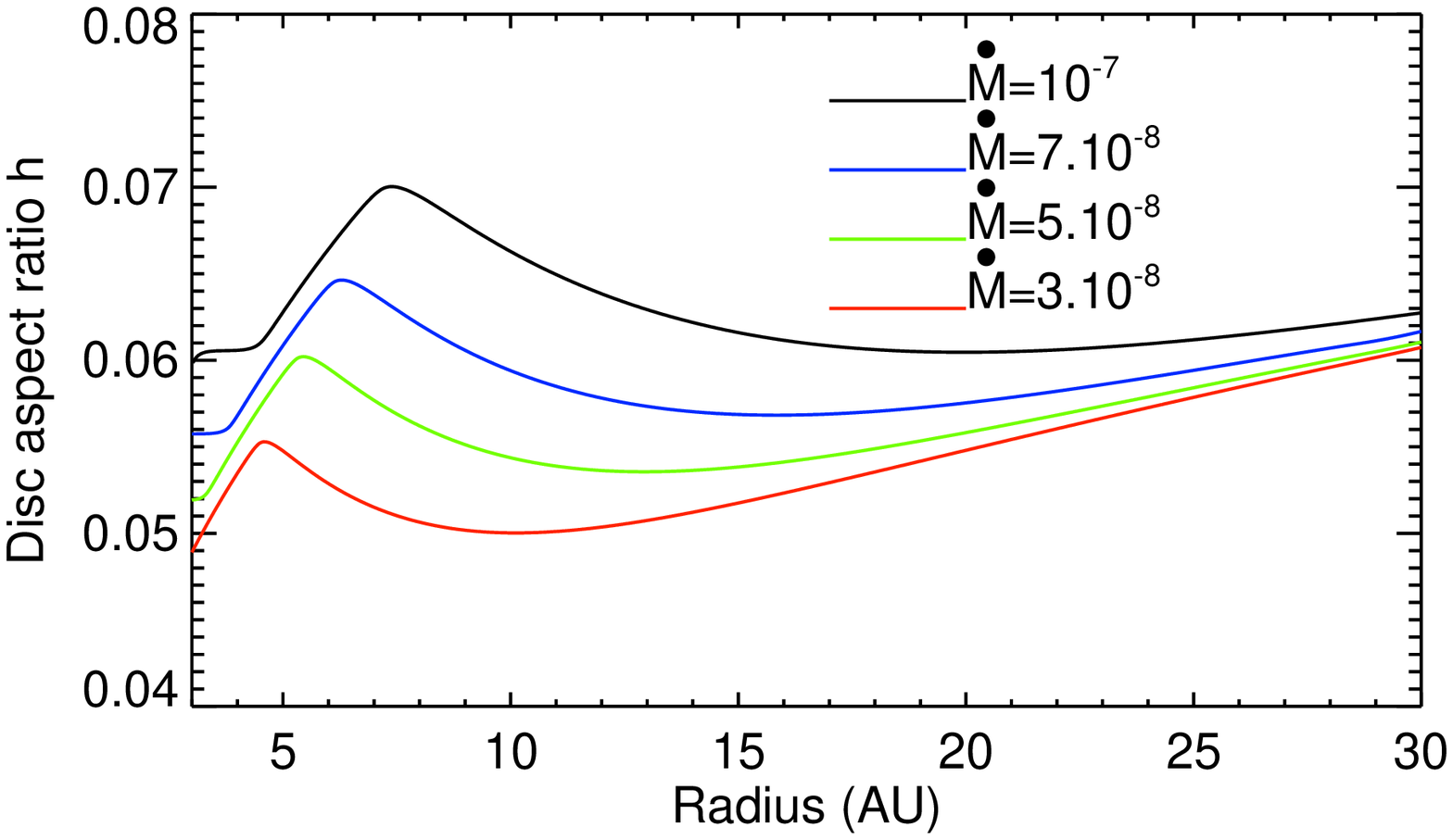}
\includegraphics[width=\columnwidth]{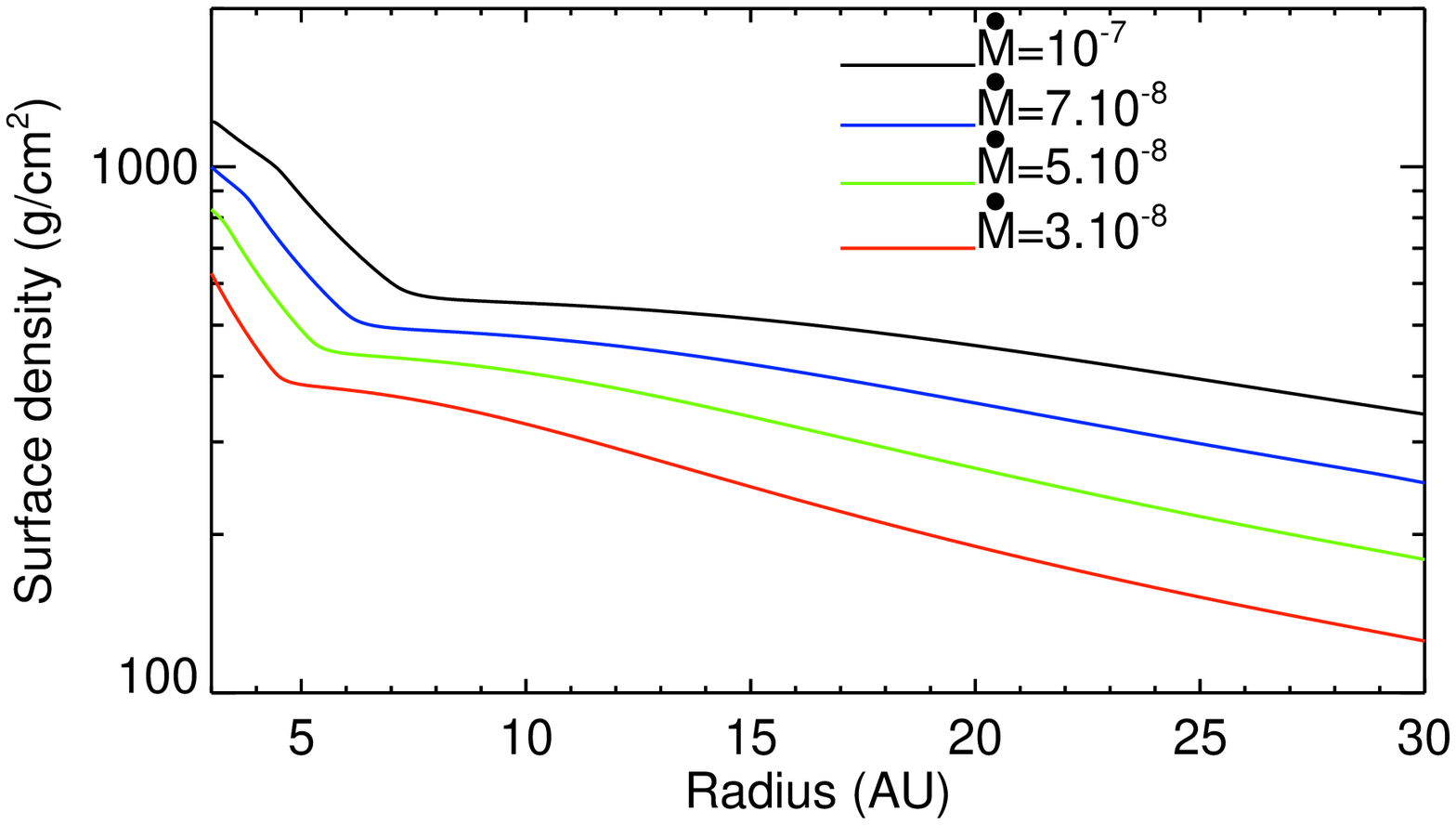}
\caption{{\it Upper panel}:  Disc aspect ratio as a function of radius for the different accreting disc models we consider.   {\it Lower panel:} Disc 
surface density for the different models.}
\label{fig:struct}
\end{figure}

\begin{figure*}
\centering
\includegraphics[width=0.48\textwidth]{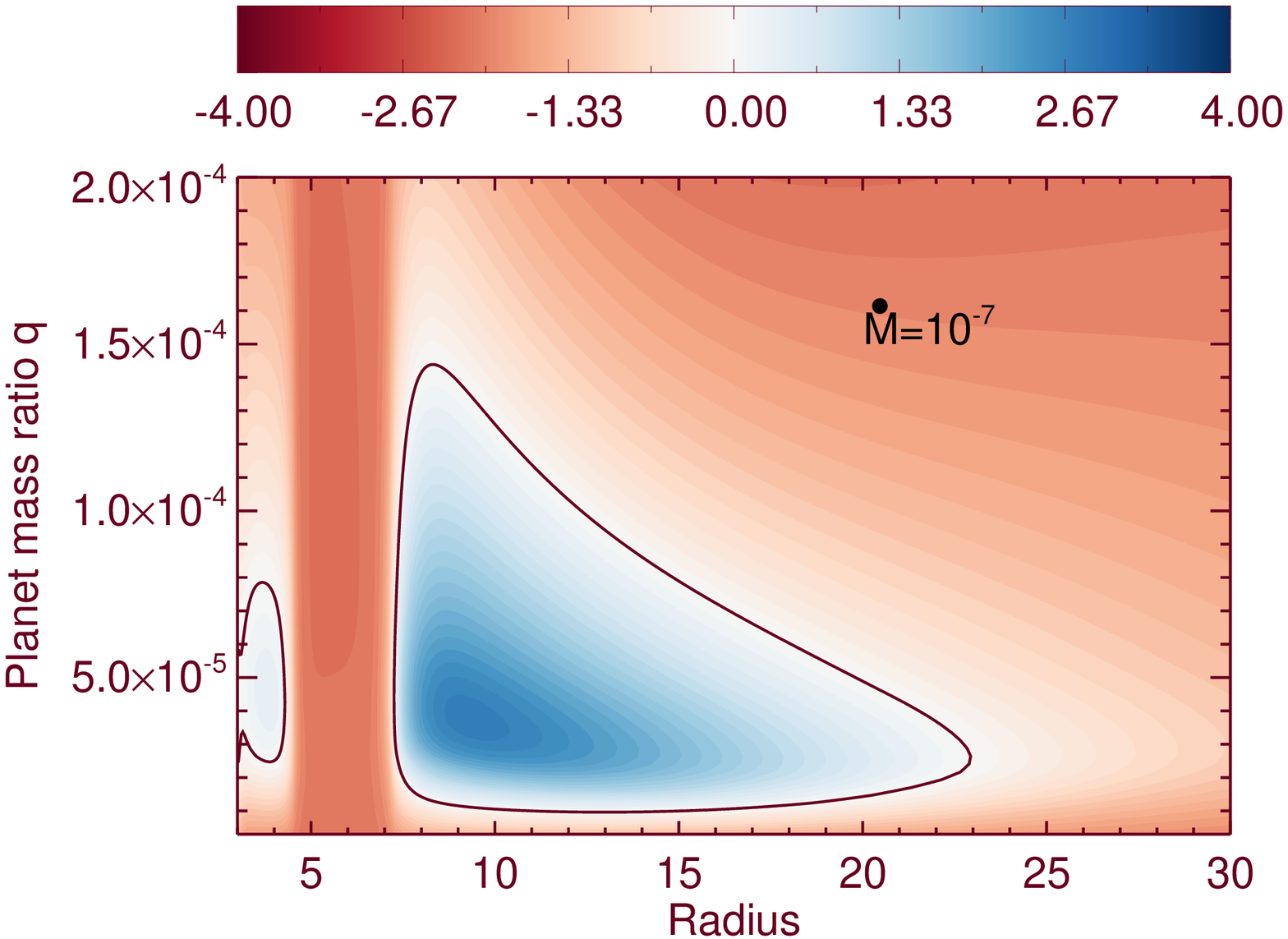}
\includegraphics[width=0.48\textwidth]{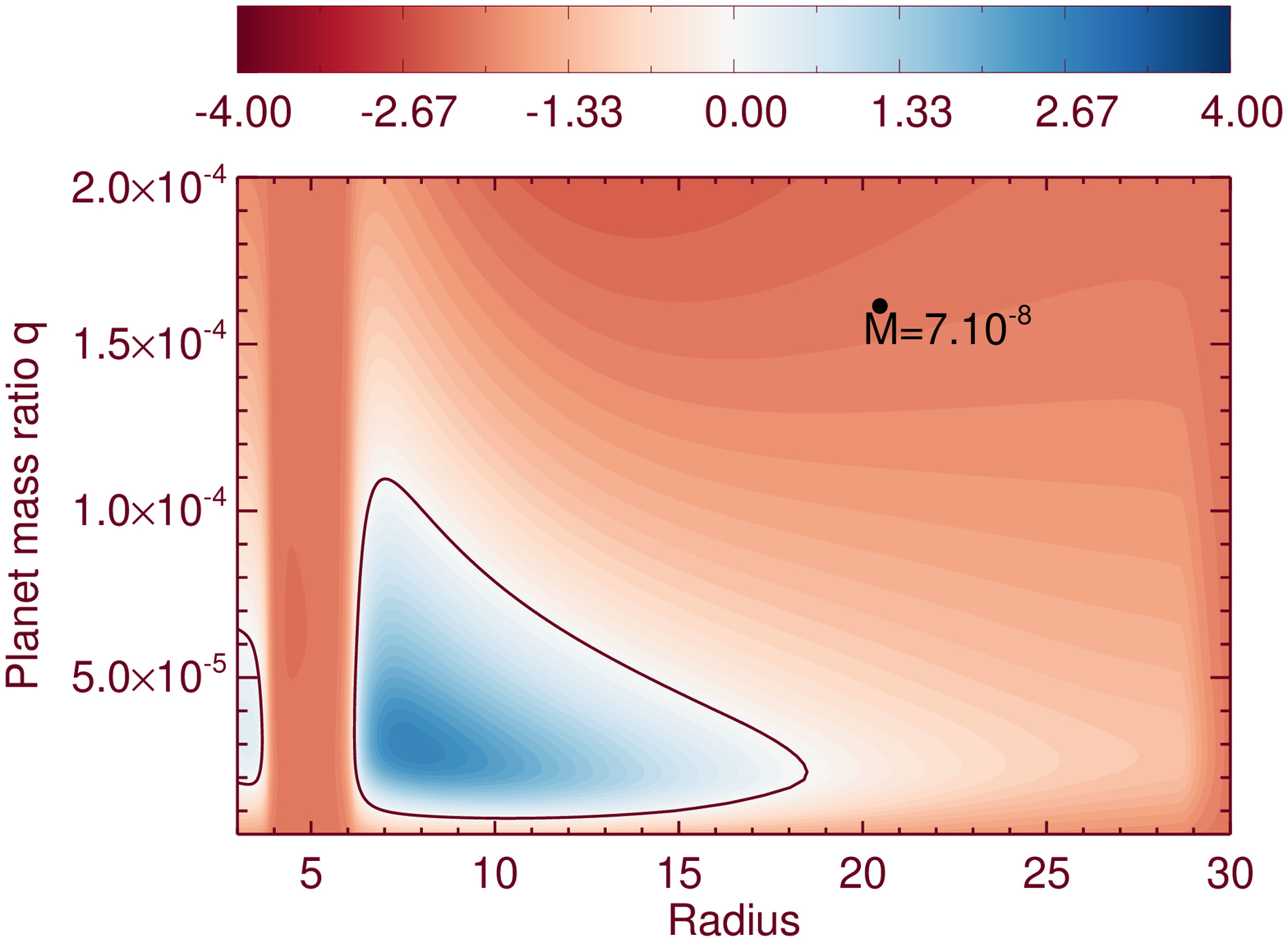}
\includegraphics[width=0.48\textwidth]{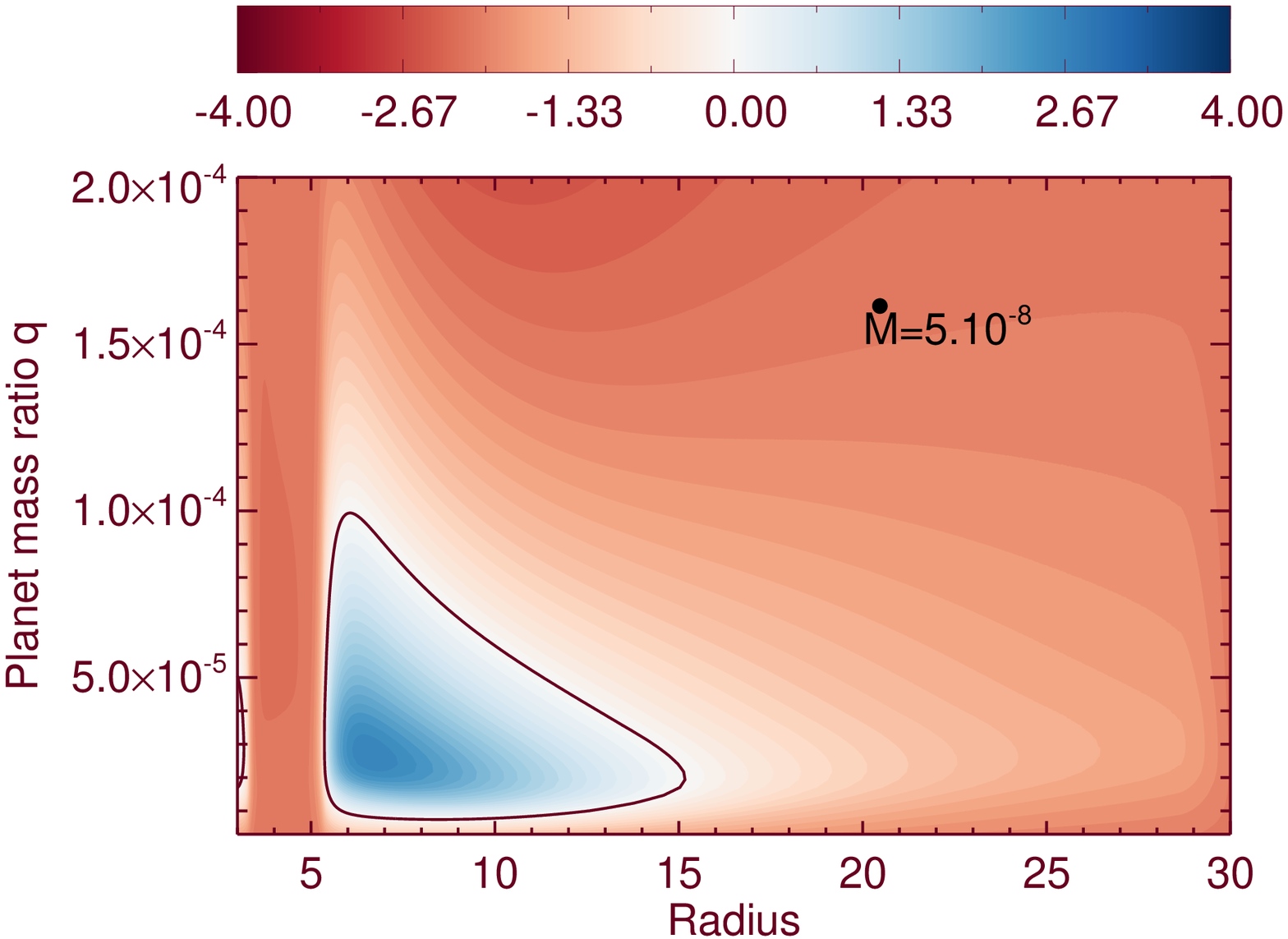}
\includegraphics[width=0.48\textwidth]{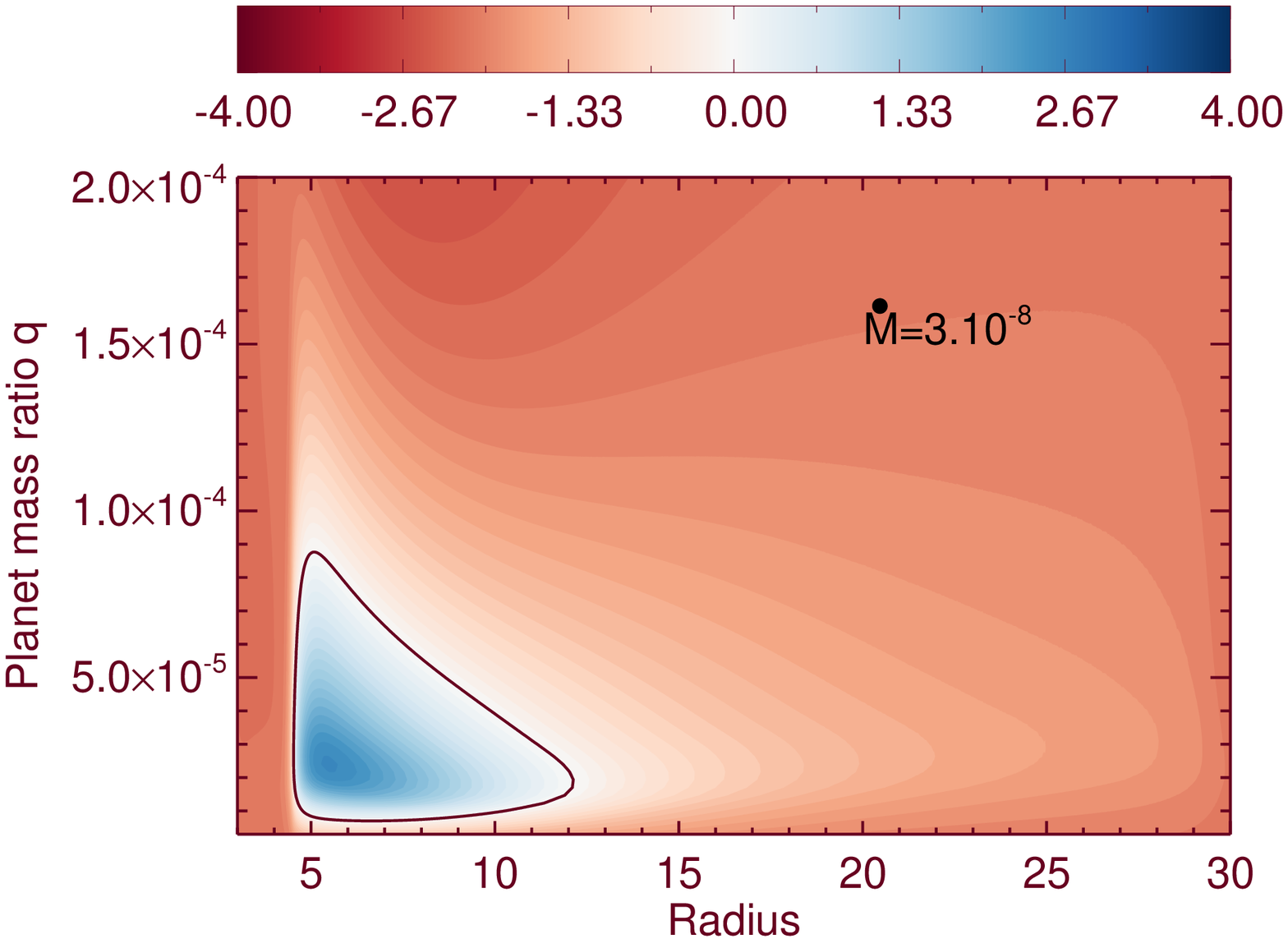}
\caption{Migration maps for the four accreting disc models we consider showing the 
predicted disc torque as a function 
of planet mass and planet radial location. The disc torque has been derived using the analytical formulae of 
Paardekooper et al. (2011) which takes into account the effect of static corotation torques only.}
\label{fig:map}
\end{figure*}

The upper panel of Fig. \ref{fig:struct} shows the disc aspect ratio as a function of radius for the different values of $\dot{M}$ we consider. 
Not surprisingly, the aspect ratio increases with $\dot{M}$ in the inner parts of the disc where viscous heating dominates over 
stellar irradiation. This  is simply a consequence of the viscous heating rate scaling linearly with $\dot{M}$. In the outer parts of the disc, however, 
stellar irradiation is the main source of heating and this leads to a flared structure with $H/R\propto R^{2/7}$ and which is almost 
independent of the value for $\dot{M}$. As expected, the transition 
from the viscous heating dominated regime to the stellar heating dominated regime occurs closer to the central star for low  $\dot{M}$. The same holds for the bump in H/R at $5-10$ AU that results from  an opacity transition.  Both the 
location and amplitude of the bump get smaller as $\dot{M}$ decreases, consistent with the results of Bitsch et al. (2014).\\
We plot in the lower panel of Fig. \ref{fig:struct} the surface density profile of the disc for each value of $\dot{M}$.  As the disc tries to  
maintain a constant accretion flow  $\dot{M}=3\pi \nu \Sigma$ with $\nu=\alpha h^2 R^2\Omega$ , changes in the disc aspect ratio  are correlated with variations in the disc surface density (Bitsch et al. 2014). Because the accretion flow is constant throughout the disc,  the strong positive gradient in $H/R$ in the very inner parts of the disc is compensated  
by a strong negative gradient in $\Sigma$.  Also,  it is clear  by comparing the upper and lower panels of Fig. \ref{fig:struct} that the bumps in $H/R$  in the inner disc correspond to jumps in 
the surface density. Regarding  the  outer flared parts of the disc, these give rise to a much shallower surface density 
since $H/R \propto R^{2/7}$ there. Again, these results are in perfect agreement with the results of Bitsch et al. (2014). \\

In Fig. \ref{fig:map} we present for each value of $\dot{M}$  the corresponding migration map,  that shows  the disc torque as a function of  planet to 
star mass ratio $q$ and orbital radius. These were calculated  using the analytical formulae of Paardekooper et al. (2011) 
for Type I migration that include 
saturation effects for the corotation torque but assume that the  planet evolves on fixed circular orbit.  Possibly strong dynamical torques 
are therefore not accounted for when computing these migration maps.  Because a positive gradient in $H/R$ tends to give rise to a positive 
entropy gradient  and consequently  to a negative corotation torque, Type I migration is directed inward in the outer parts of the disc where stellar irradiation is the main source of heating. In the inner parts of the 
disc where viscous heating dominates, however,  migration can be directed outward due to a positive corotation torque resulting from a negative 
entropy gradient.  Looking at the lower panel of Fig. \ref{fig:struct}, it is clear that  assuming $\sigma \propto R^{-\sigma}$,  the surface density index $-\sigma$  in the region where outward migration 
can occur is such that $\sigma < 1.5$, resulting in a positive vorticity-related horseshoe drag. However, we find that  this component of the corotation torque 
is typically smaller than the one related to the entropy gradient by a factor of $\sim 3$, which is therefore the main engine for driving outward migration.
Of course, outward migration can only occur provided that the corotation torque is close to its optimal value,  which is reached when the viscous/thermal  timescale  
$t_d$ across 
the  horseshoe region  is approximately equal to half the libration timescale $t_{lib}$ (Baruteau \& Masset 2013). At thermodynamical equilibrium,  $t_d\sim x_s^2/\nu$ where $x_s\sim 1.2 a(q/h_p)^{1/2}$ (Masset et al. 2006) denotes the half-width of the horseshoe region, with $a$  the planet semi major axis and 
$h_p$ the disc aspect ratio at the planet location. Given that $t_{lib}=8\pi a /3\Omega_p x_s$, where $\Omega_p$ is the planet 
angular velocity, it is straightforward 
to show that planets that can 
experience outward migration have planet-to-star mass ratios $q$ such that:
\begin{equation}
q\sim 2.6 \alpha^{2/3} h_p^{7/3}
\label{eq:q}
\end{equation}
For $\alpha=2\times 10^{-3}$ and $h_p=0.06$  this gives $q\sim 6\times10^{-5}$, consistent with  the migration maps shown in Fig. \ref{fig:map}. Because 
the corotation torque for planets with mass ratios given by Eq. \ref{eq:q} is close to its fully unsaturated value, these are good candidates to experience 
dynamical torques provided the disc is massive enough. For the disc models we consider here, it is also interesting to note that this mass 
range  corresponds to the typical mass range for runaway gas accretion in the classical core accretion 
scenario (Pollack et al. 1996). We will discuss the effect of gas accretion on dynamical torques in Sect. \ref{sec:accretion}.\\

\begin{figure}
\centering
\includegraphics[width=\columnwidth]{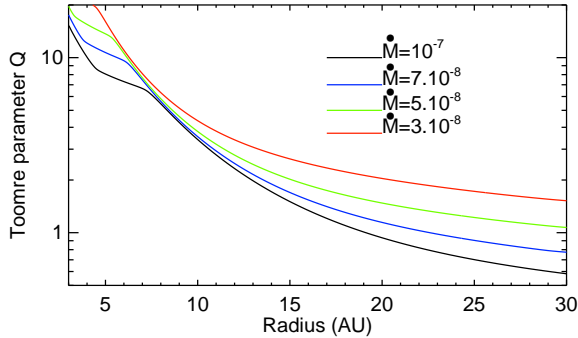}
\caption{Toomre stability parameter as a function of radius for the different accreting disc models}
\label{fig:toomre}
\end{figure}

  In a previous study (Pierens 2015), we showed that  a necessary condition for dynamical torques to come into action is that 
the Toomre parameter at the planet location ${\cal Q}=h_p/\pi \Sigma_p a^2$ satisfies the following relation:
 
 \begin{equation}
 {\cal Q}<\frac{3.7}{\gamma}(\gamma \Gamma/\Gamma_0)
 \label{eq:calq}
 \end{equation}
 
where $\gamma \Gamma/\Gamma_0$ is the dimensionless torque that takes 
into account possible saturation effects of the corotation torque, and $\Gamma_0=(q/h_p)^2\Sigma_p a \Omega_p$, with 
$\Sigma_p$ the disc surface density at the planet location.  When the condition given by Eq. \ref{eq:calq} is met during the course of migration, the drift 
timescale across the horseshoe region becomes shorter than the libration timescale, with the implication that the horseshoe 
region contracts into a tadpole-like region. Fig. \ref{fig:toomre} shows the Toomre parameter 
${\cal Q}$ as a function of radius for each value of $\dot{M}$ we consider. As expected, ${\cal Q}$ decreases with radius and can reach values 
below the gravitational stability limit for our most massive disc, with accretion rates ${\dot M} \ge 7\times 10^{-8}\;M_\odot/yr$. Comparison with Fig. \ref{fig:struct}, however, indicates that the 
outward-migrating region that we will focus on in this study is gravitationally stable for all values of ${\dot M}$.

From the 
above discussion, we expect the effect  of dynamical torques to be stronger for large values of ${\dot M}$, i.e., for more massive discs.  This means that the influence of dynamical torques should be most pronounced early in a disc's evolution. There are reasons for this: i) ${\cal Q}$ 
decreases as  ${\dot M}$ increases in the outer regions of our accretion disc models,  and ii)  the outward-migrating region extends outward 
as ${\dot M}$ increases (Bitsch et al. 2014), so that an outward migrating protoplanet is more prone to reaching disc regions where the constraint given 
by Eq. \ref{eq:calq} is satisfied if ${\dot M}$ is high enough.

\section{Dynamical torques on giant planet cores}
\label{sec:results}
\begin{figure}
\centering
\includegraphics[width=\columnwidth]{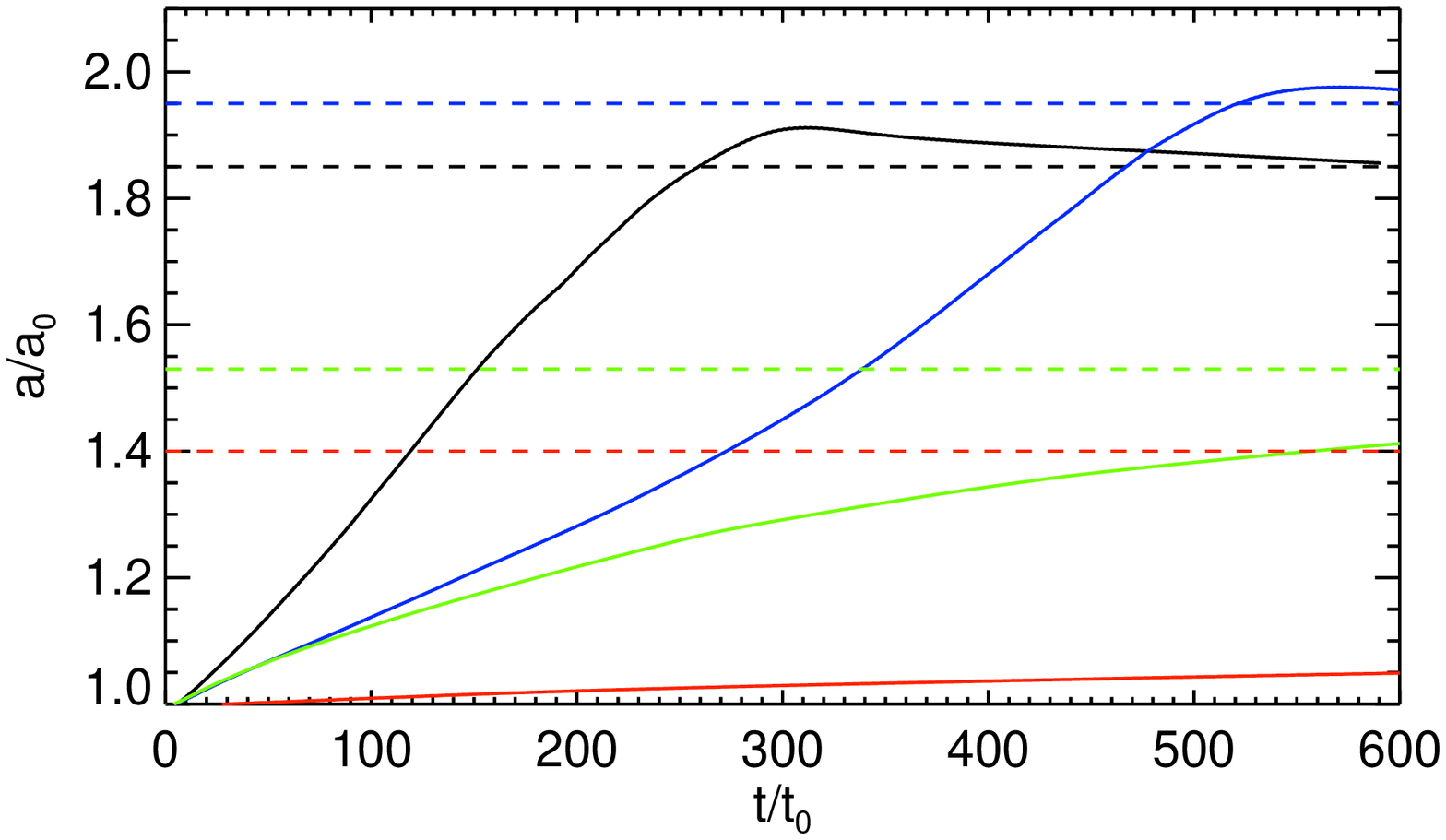}
\includegraphics[width=\columnwidth]{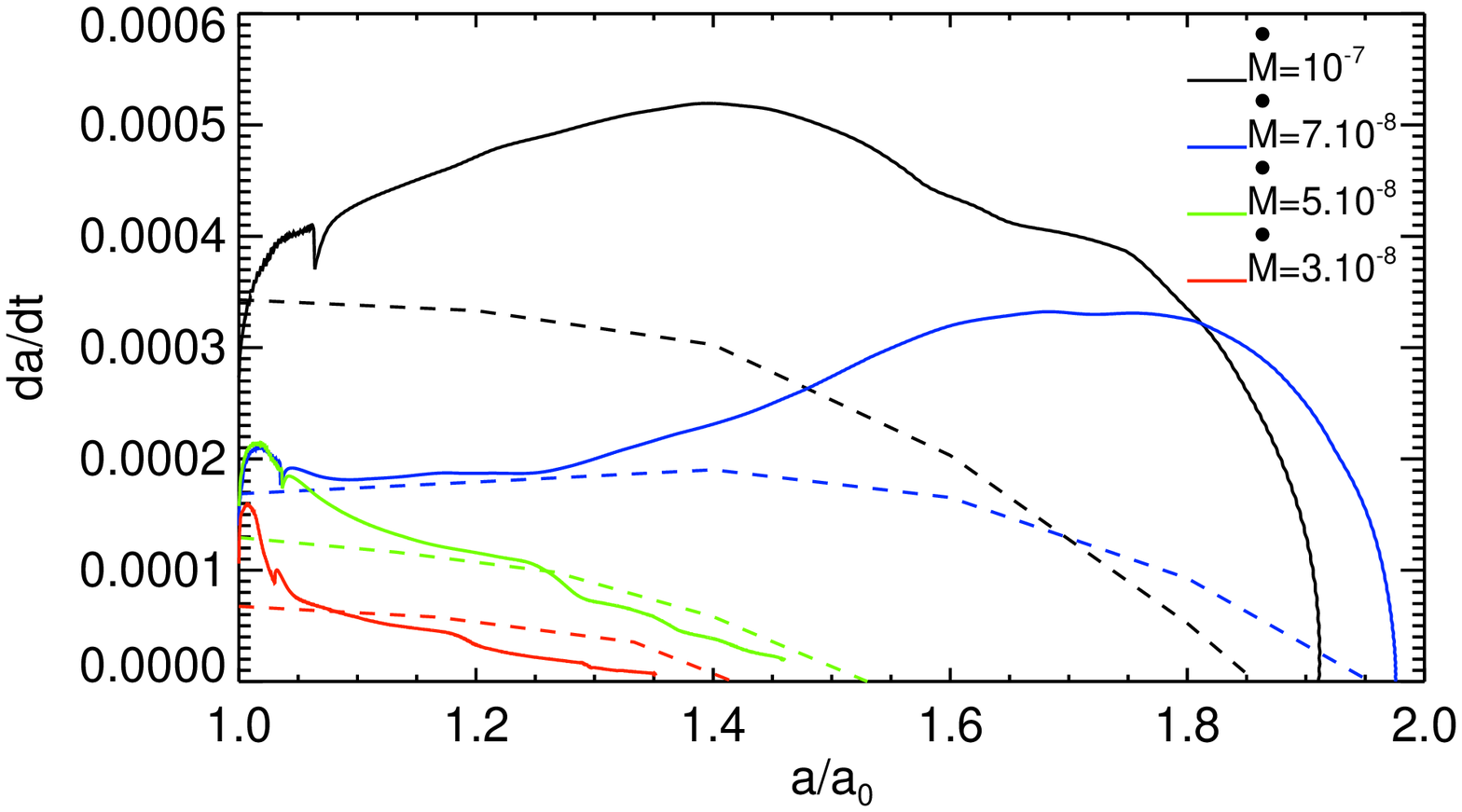}
\caption{{\it Upper panel}: Time evolution of the semi major axis of a $20M_\oplus$ 
protoplanet relative to its initial location $a_0$, for the different values of ${\dot M}$ we consider. The dashed line shows the location 
of the zero-torque where the saturated corotation torque and the Lindblad torque balance each other, as predicted by static torques.  Time is measured in units of the orbital period at $a_0$. {\it Lower panel:} corresponding drift rates as a function of normalized semi major axis. The dashed line shows the drift rates that are expected 
from static torques only. }
\label{fig:a20}
\end{figure}

\begin{figure}
\centering
\includegraphics[width=\columnwidth]{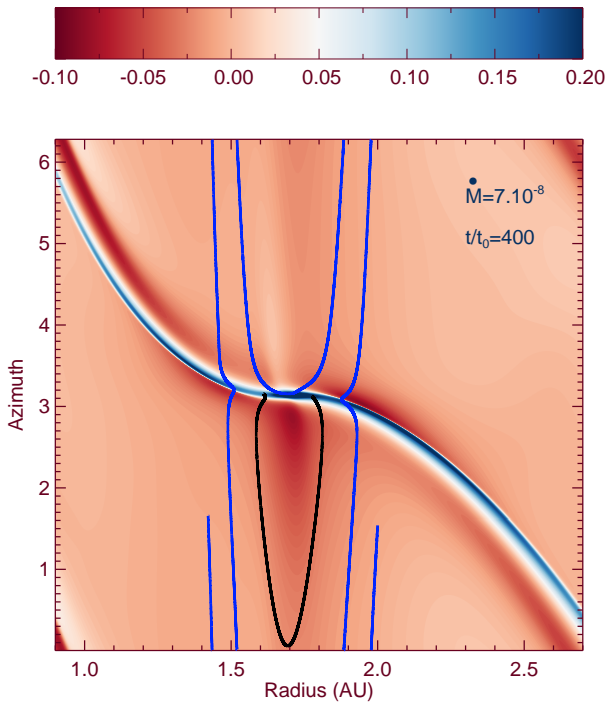}
\caption{Contours of the perturbed surface density near a $20M_\oplus$ protoplanet for the 
accreting  disc model with ${\dot M}=7\times 10^{-8}\;M_\odot/yr$ at $t=400 t_0$, where $t_0$ refers 
to the orbital period at the initial planet location. Overplotted as black line is a streamline that corresponds 
to the librating material that moves with the planet as it migrates. Overplotted as blue line is a streamline 
that corresponds to material that flows across the planet orbit as it migrates.}
\label{fig:2d}
\end{figure}

\begin{figure}
\centering
\includegraphics[width=\columnwidth]{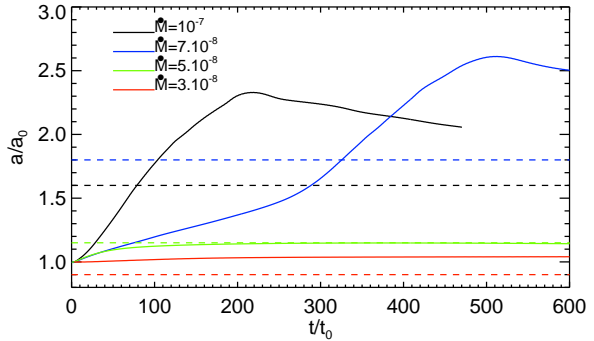}
\caption{Time evolution of the semi major axis of a $30M_\oplus$ 
protoplanet relative to its initial location $a_0$, for the different values of ${\dot M}$ we consider.  The dashed line shows the location 
of the zero-torque where the saturated corotation torque and the Lindblad torque balance each other, as predicted by static torques. Time is measured in units of the orbital period at $a_0$.}
\label{fig:a30}
\end{figure}

In this section we provide  a quantitative 
analysis of the dependence of dynamical torques experienced by protoplanets on ${\dot M}$, with the goal of estimating the critical accretion rate 
above which dynamical torques strongly affect the orbital evolution of protoplanets.  We focus on  protoplanets 
with masses of 20-30 $M_\oplus$,  since these are more sensitive to dynamical torques due to their  corotation torque being 
close to their fully unsaturated value.  We neglect gas accretion in this section but will consider this process  in the next section. \\
 These planets are released immediately after insertion in the disc and the initial planetary orbital radius $a_0$ is located in the region where outward migration is expected to occur from the migration maps in Fig. \ref{fig:map}. It  takes different 
values that  depend on the value for ${\dot M}$ that is considered. It is set to $a_0=11$ AU for ${\dot M}=10^{-7}\;M_\odot/yr$, $a_0=8$ AU for 
${\dot M}=5\times10^{-8}$,  $7\times 10^{-8}\;M_\odot/yr$ and  $a_0=6$ AU for ${\dot M}=3\times10^{-8}\;M_\odot/yr$. For each ${\dot M}$, we plot in the upper panel of 
Fig. \ref{fig:a20} the evolution of the semimajor axis relative to $a_0$  of a $20$ $M_\oplus$ planet as a function of time (measured in units of planet orbits $t_0$ at $a_0$).  The corresponding  drift rates versus normalized semi major axis are shown in the lower panel of  Fig. \ref{fig:a20}, and
compared with the migration rate expected from static torques only. The  migration rate assuming a static torque has  been simply obtained by measuring, at different 
orbital radii, the torque experienced by a $20$ $M_\oplus$ planet evolving on a fixed circular orbit. We see that for ${\dot M}\le 5\times10^{-8}\;M_\odot/yr$,  there is decent agreement between the actual migration rates of the planet and those derived assuming a static torque. For 
${\dot M}\ge 7\times10^{-8}\;M_\odot/yr$, however, 
these two estimations can significantly differ from each other, with the actual planet drift rates being larger than those obtained from static torques 
by a factor of $\sim 2$ typically. For ${\dot M}= 7\times10^{-8}\;M_\odot/yr$, both estimates  show almost similar values initially, but these diverge at later times. This is a clear indication that in this case, dynamical torques can play an important role in the planet's orbital evolution. As the 
planet migrates outward, the Toomre parameter at the planet location continuously decreases until a point in time where the condition given 
by Eq. \ref{eq:calq} becomes eventually  satisfied, resulting in the planet horseshoe region to no longer extend to the full $2\pi$ in azimuth.  
This is  illustrated  in Fig. \ref{fig:2d}, which shows  contours of
  the perturbed surface density at $t=400t_0$ (as well as a few streamlines) for the case with ${\dot M}= 7\times 10^{-8}\;M_\odot/yr$. These clearly show that the horseshoe has contracted into a tadpole-like region (dark streamline). Since this region moves with the planet as it migrates, it has a negative feedback on migration, whereas disc material flowing across the planet orbit (represented as blue streamline) has a positive feedback on migration. From Fig.   \ref{fig:2d}, 
  however,  it is clear that the 
  tadpole-like region is underdense relative to the ambient disc, which arises because the planet migrates in the direction set by the 
  entropy-related corotation torque. As discussed in Pierens (2015), this results in  a net positive feedback on migration and the possibility for the planet 
  to reach high drift rates, consistently with what is observed in the lower panel of Fig. \ref{fig:a20}. \\
  The orbital evolution of $30$ $M_\oplus$ planets embedded in our four accreting discs is presented in Fig. \ref{fig:a30}. Again, we find strong 
  dynamical torques provided that ${\dot M} \ge 7\times10^{-8}\;M_\odot/yr$. For ${\dot M} \le 5\times10^{-8}\;M_\odot/yr$, migration is stopped close 
  to  the zero-torque radius where the 
  differential Lindblad torque counterbalances the saturated corotation torque. For higher accretion rates, however, the planet is observed to migrate beyond  the zero-torque radius and  the  transition between the viscous heating dominated and  stellar 
  heating dominated regimes. In the run with ${\dot M}=7\times10^{-8}\;M_\odot/yr$ for example, outward migration stalls at $a/a_0\sim 2.6$ 
  ($a\sim21$ AU) whereas 
  static torques  predict inward migration for $a/a_0\gtrsim 1.8$ ($a\gtrsim  14$ AU). Moreover, the transition  between the viscous heating dominated and  stellar 
  heating dominated regimes is located at  $a/a_0\sim 2.4$ ($a\sim 19$ AU) for this model. For ${\dot M}=10^{-7}\;M_\odot/yr$, outward migration is expected  to stop at 
  $a/a_0\sim 1.8$ ($a\sim 14$AU)  whereas dynamical torques make the outermost location that can be reached by the  planet be $a/a_0\sim 2.3$ 
  ($a\sim 26$ AU). Again, this is outside the transition  between the viscous heating dominated and stellar 
  heating dominated regimes which is located at $a/a_0\sim 2$ ($a\sim23$ AU) for this run.

\section{Effect of gas accretion}
\label{sec:accretion}

\begin{figure*}
\centering
\includegraphics[width=0.45\textwidth]{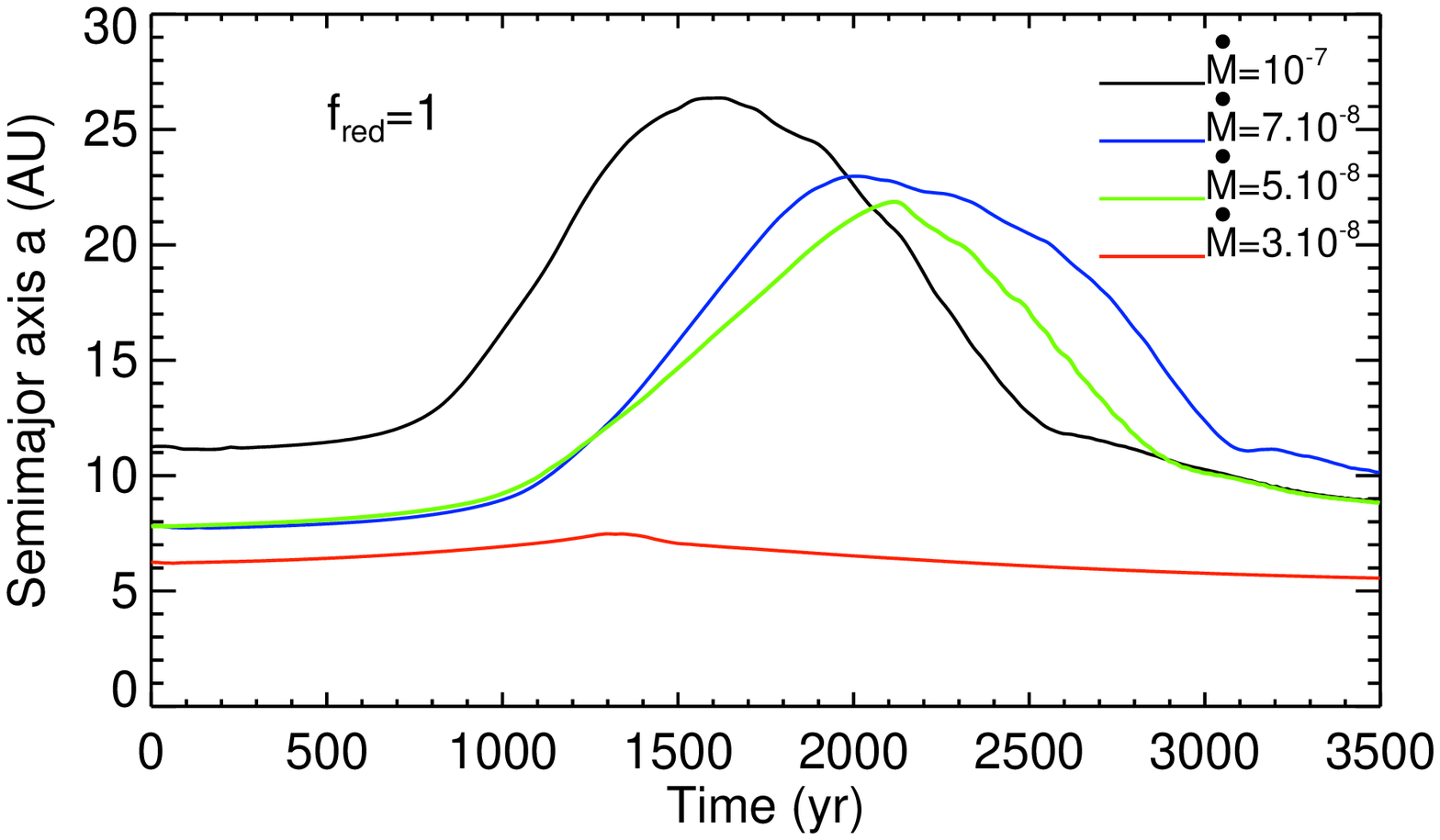}
\includegraphics[width=0.45\textwidth]{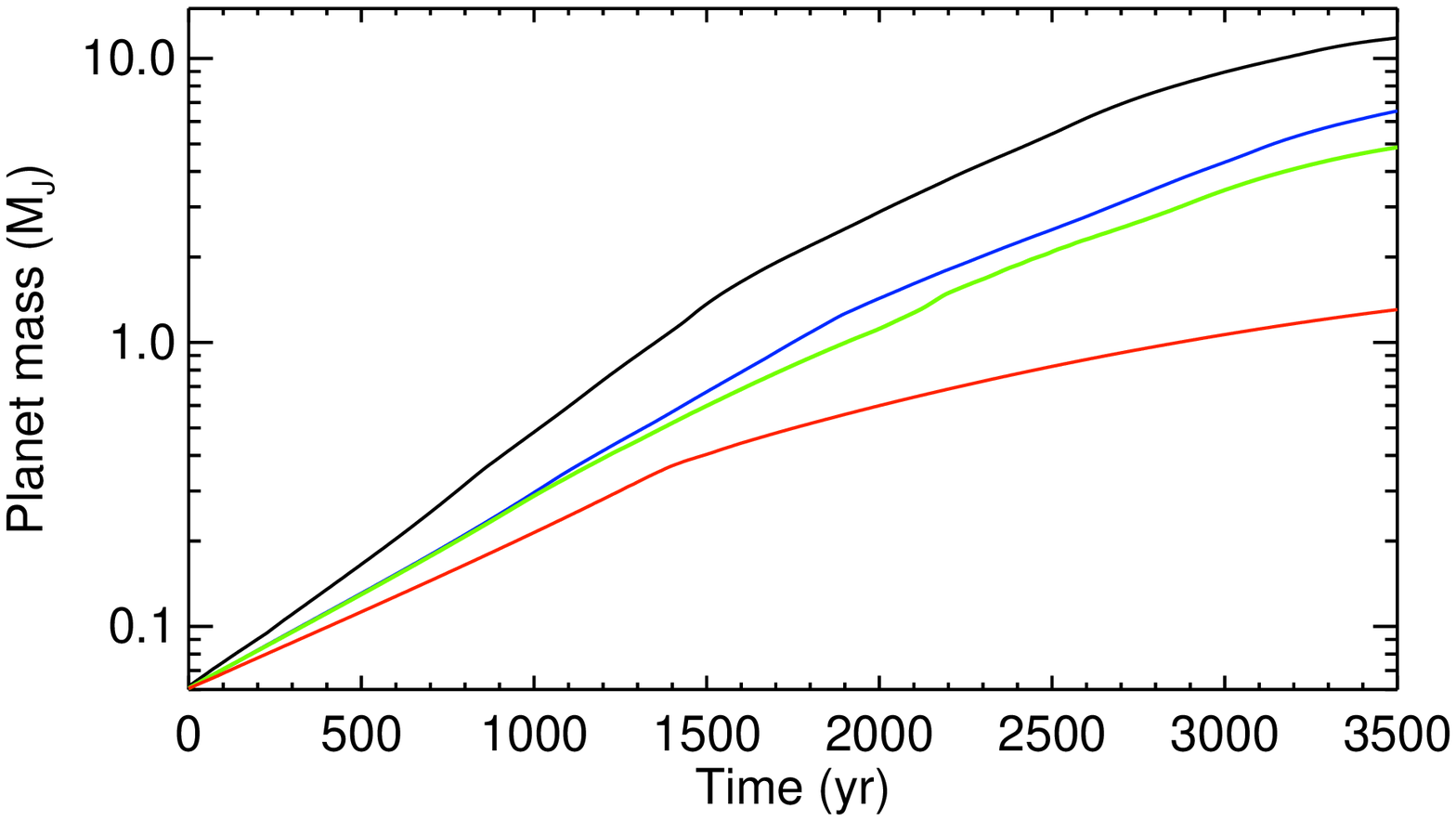}
\includegraphics[width=0.45\textwidth]{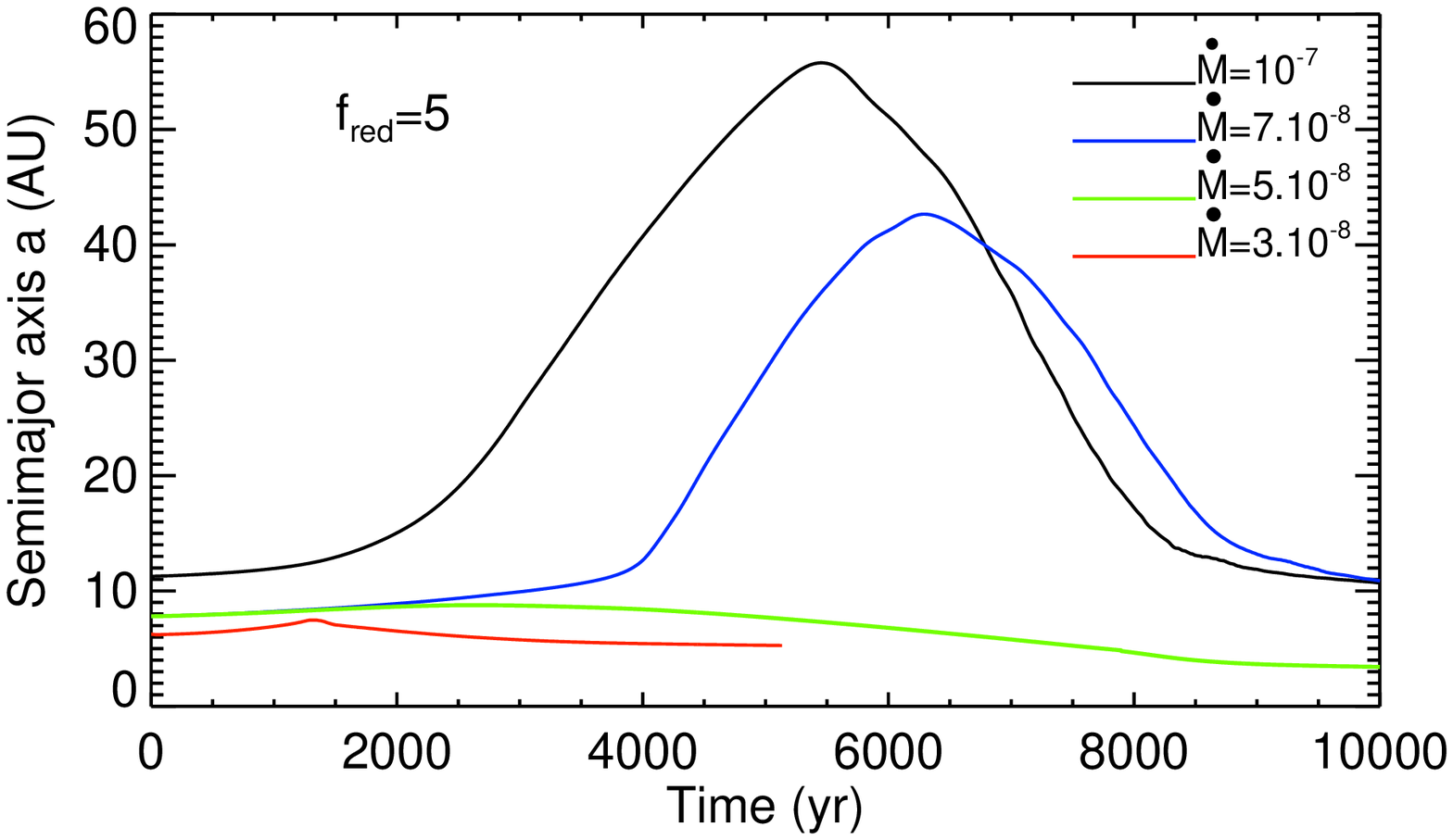}
\includegraphics[width=0.45\textwidth]{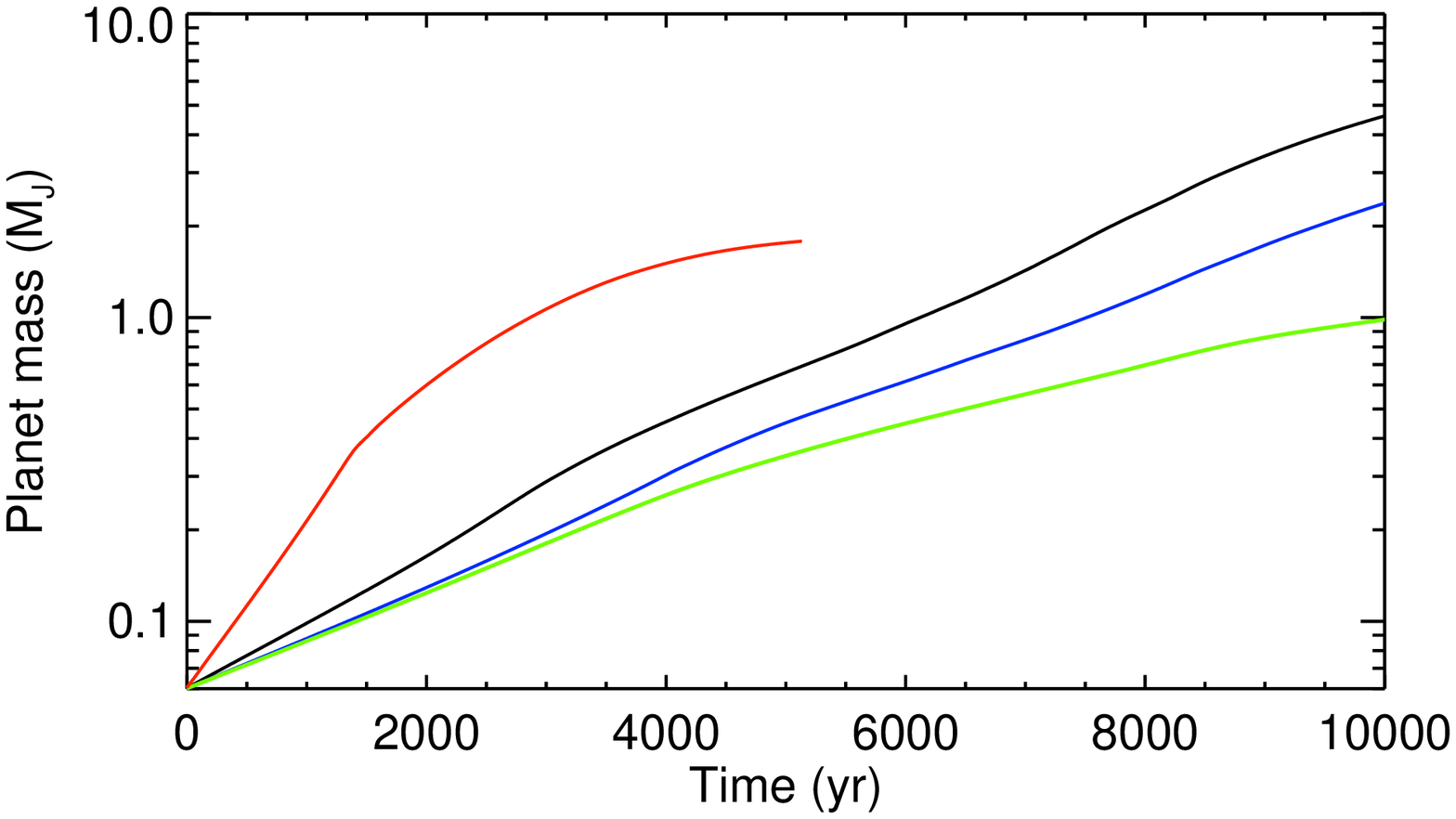}
\includegraphics[width=0.45\textwidth]{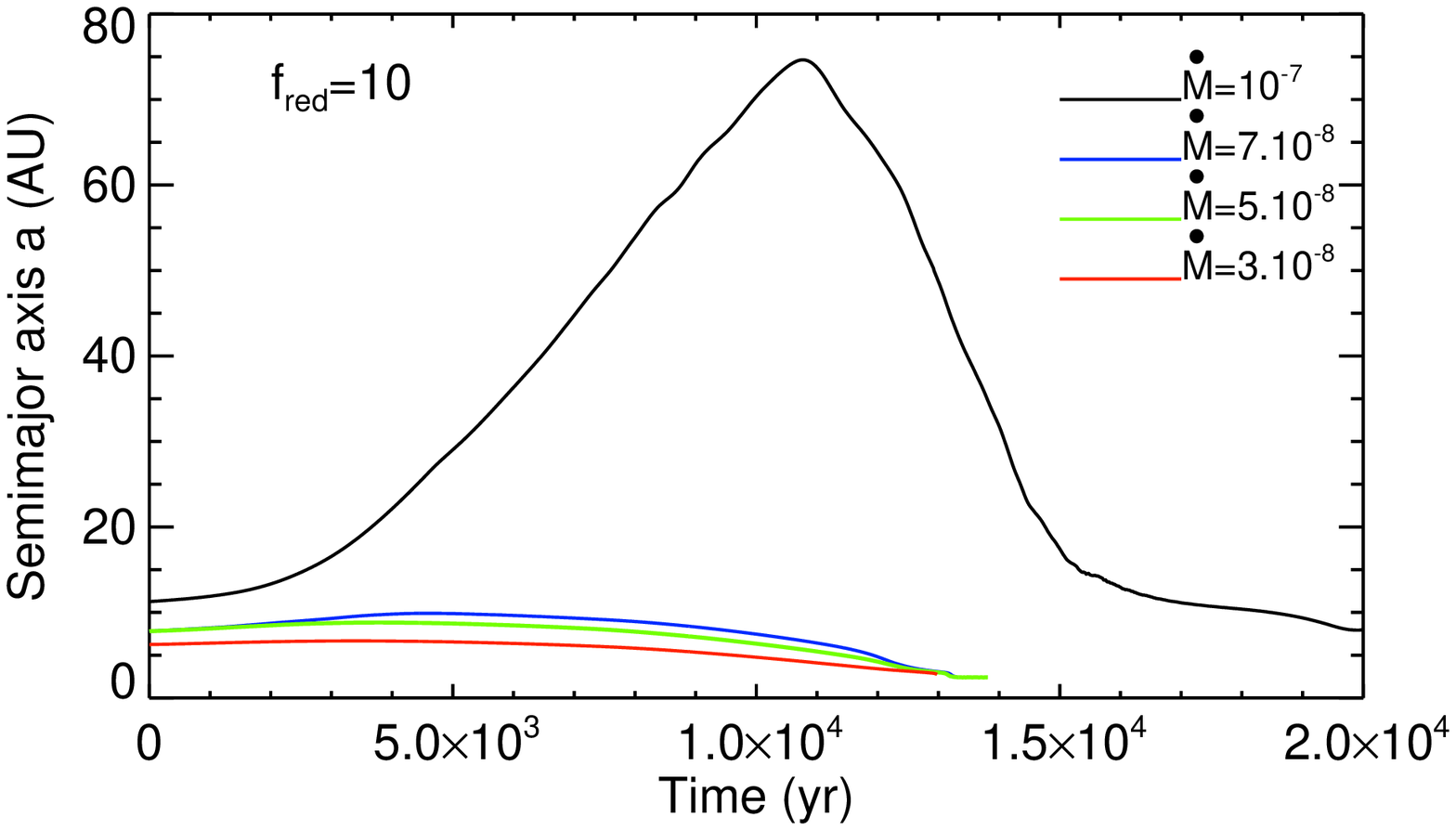}
\includegraphics[width=0.45\textwidth]{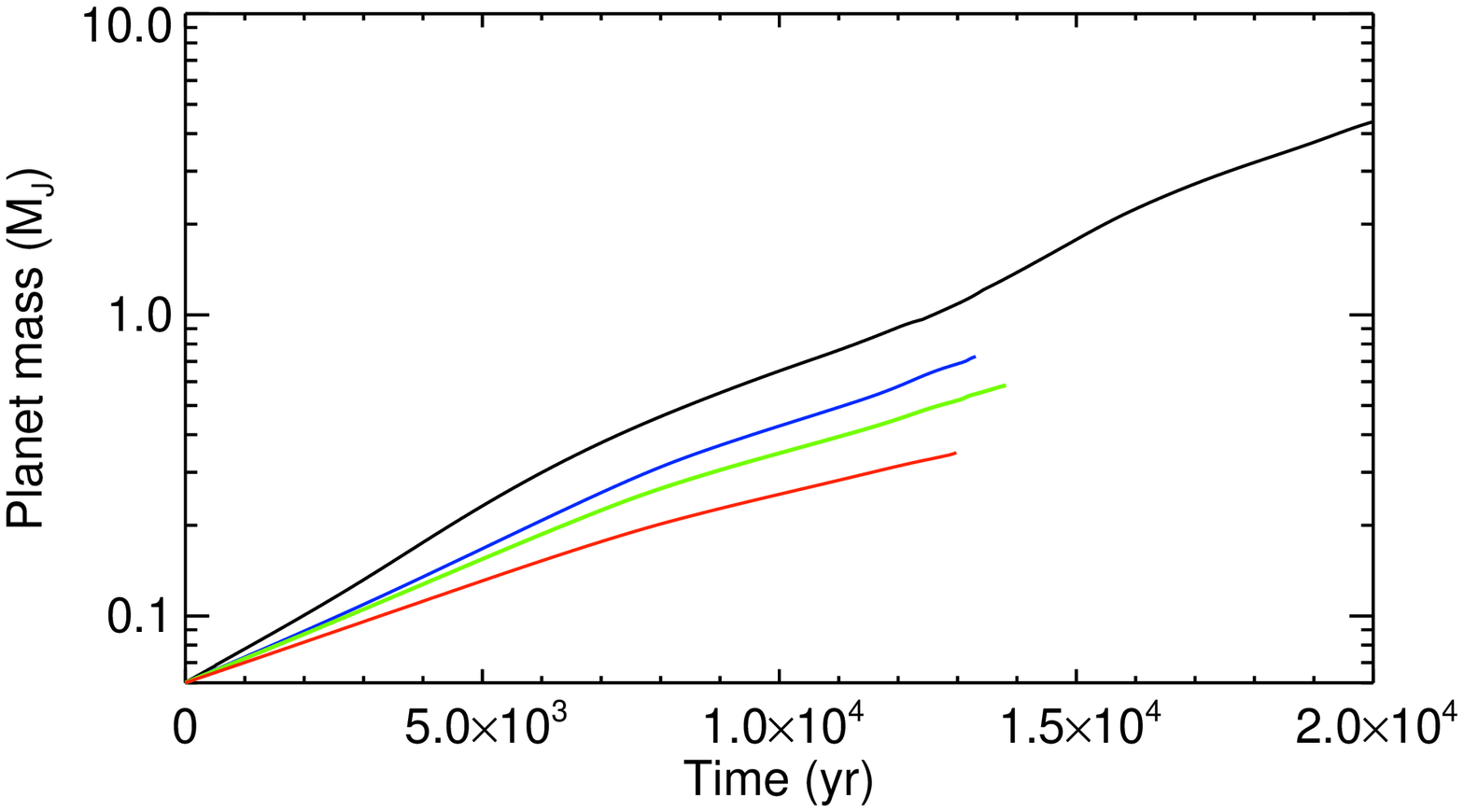}
\caption{Time evolution of the planet semi major axis (left panel) and planet mass (right panel) for the simulation
with ${\dot M}=7\times 10^{-8}\;M_\odot/yr$ and 
in which a protoplanet with initial mass of $20M_\oplus$ accretes gas material on characteristic timescale 
$t_{acc}=f_{red}t_{dyn}$.}
\label{fig:accret}
\end{figure*}

\begin{figure*}
\centering
\includegraphics[width=0.45\textwidth]{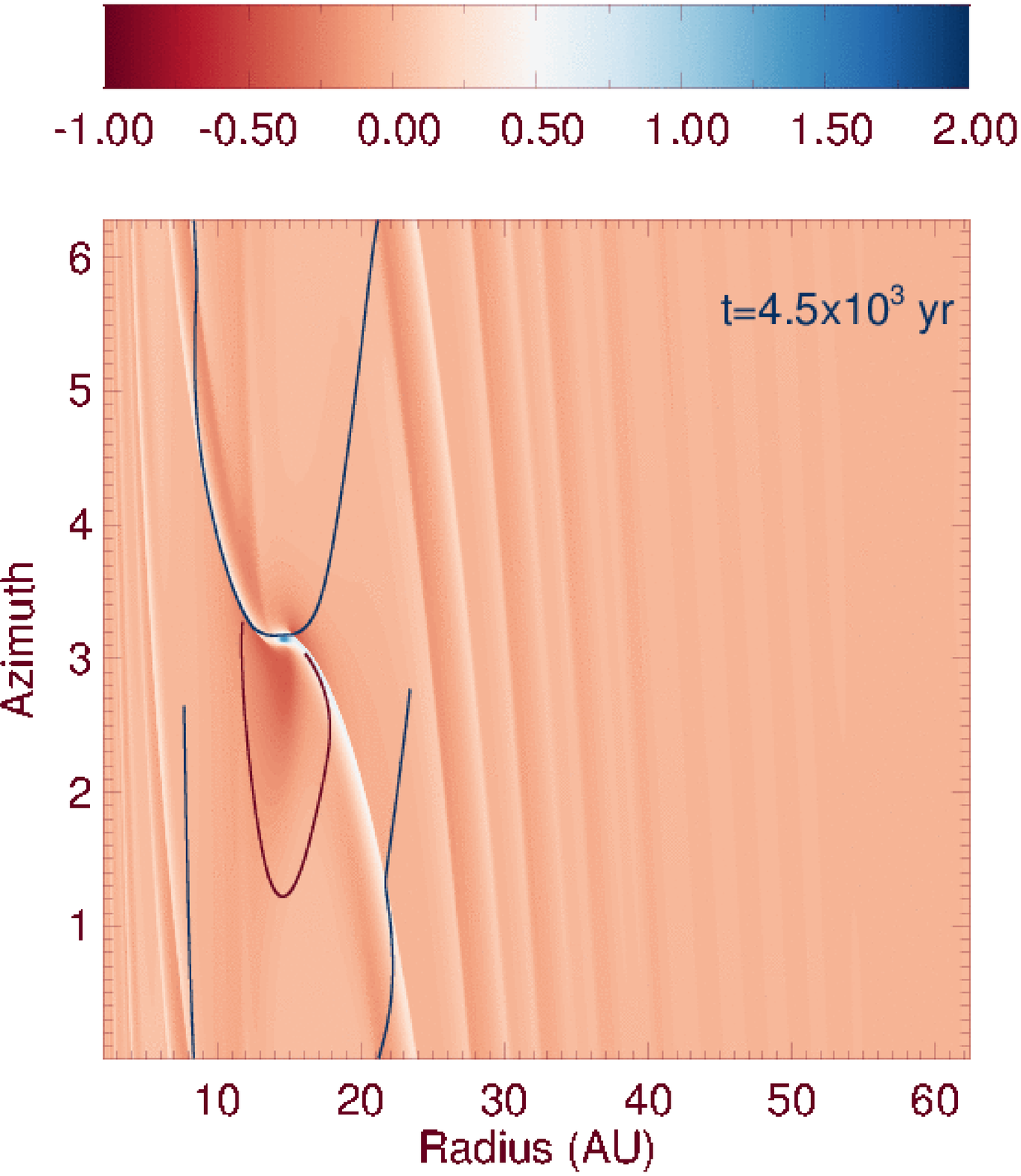}
\includegraphics[width=0.45\textwidth]{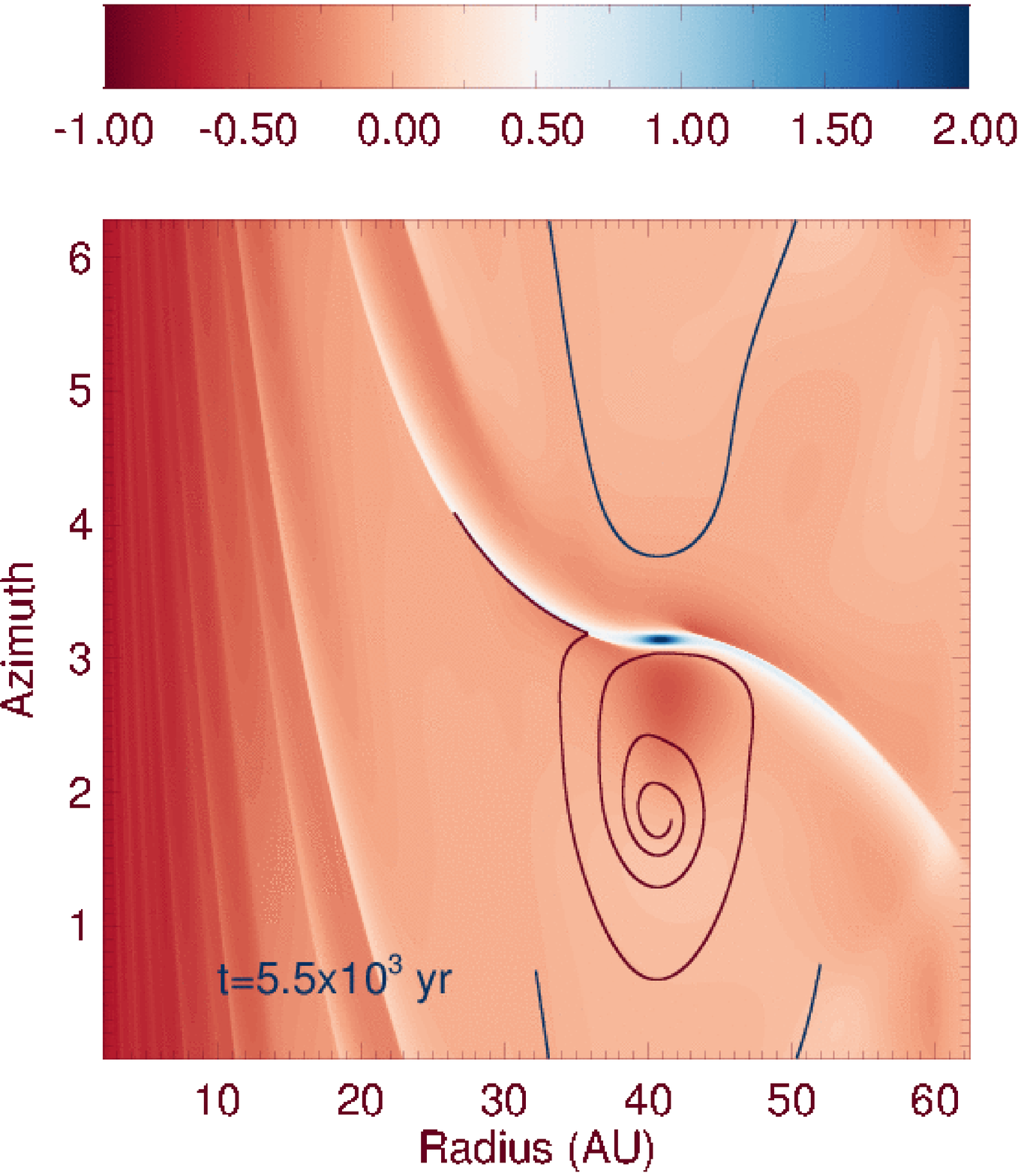}
\includegraphics[width=0.45\textwidth]{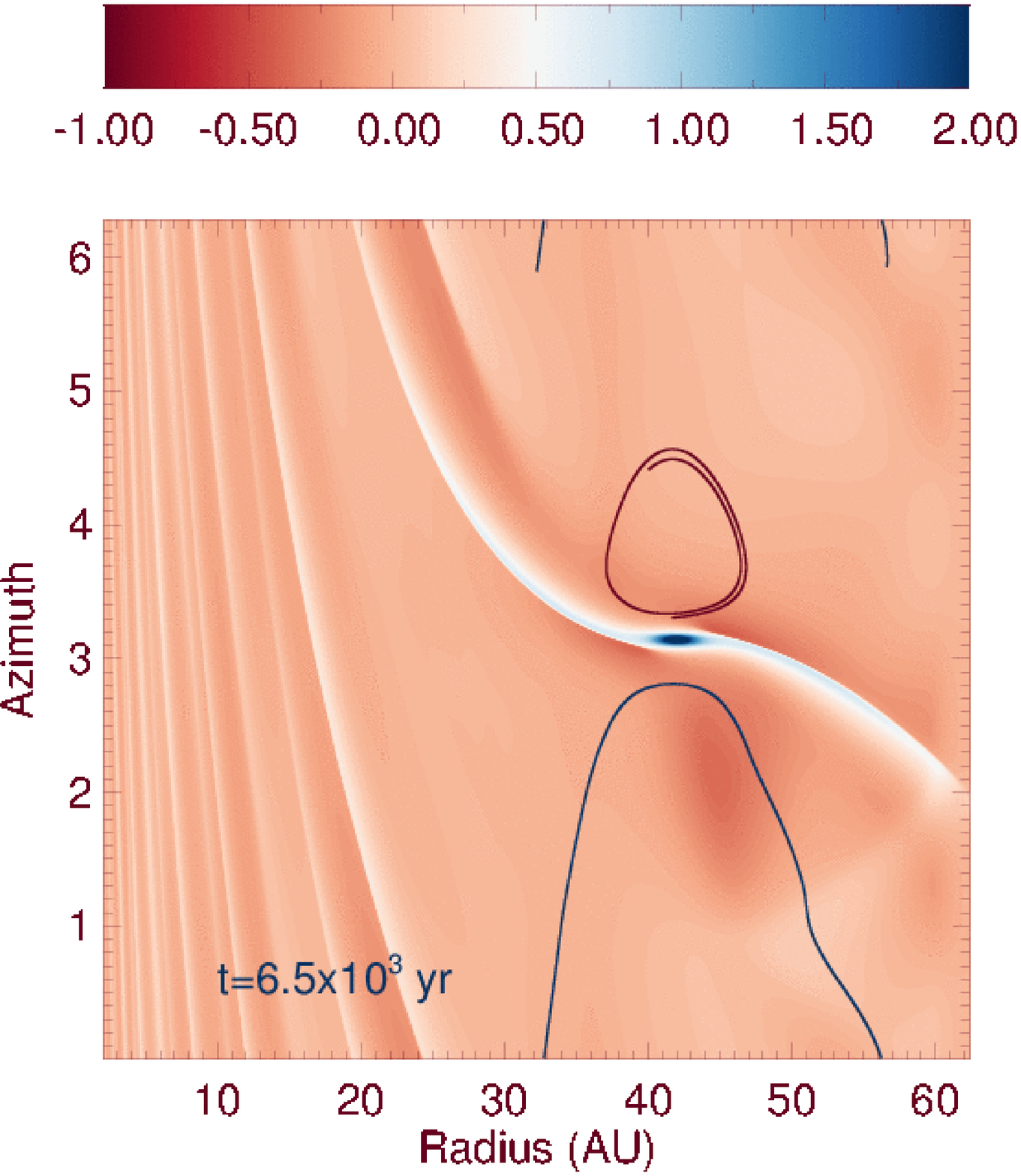}
\includegraphics[width=0.45\textwidth]{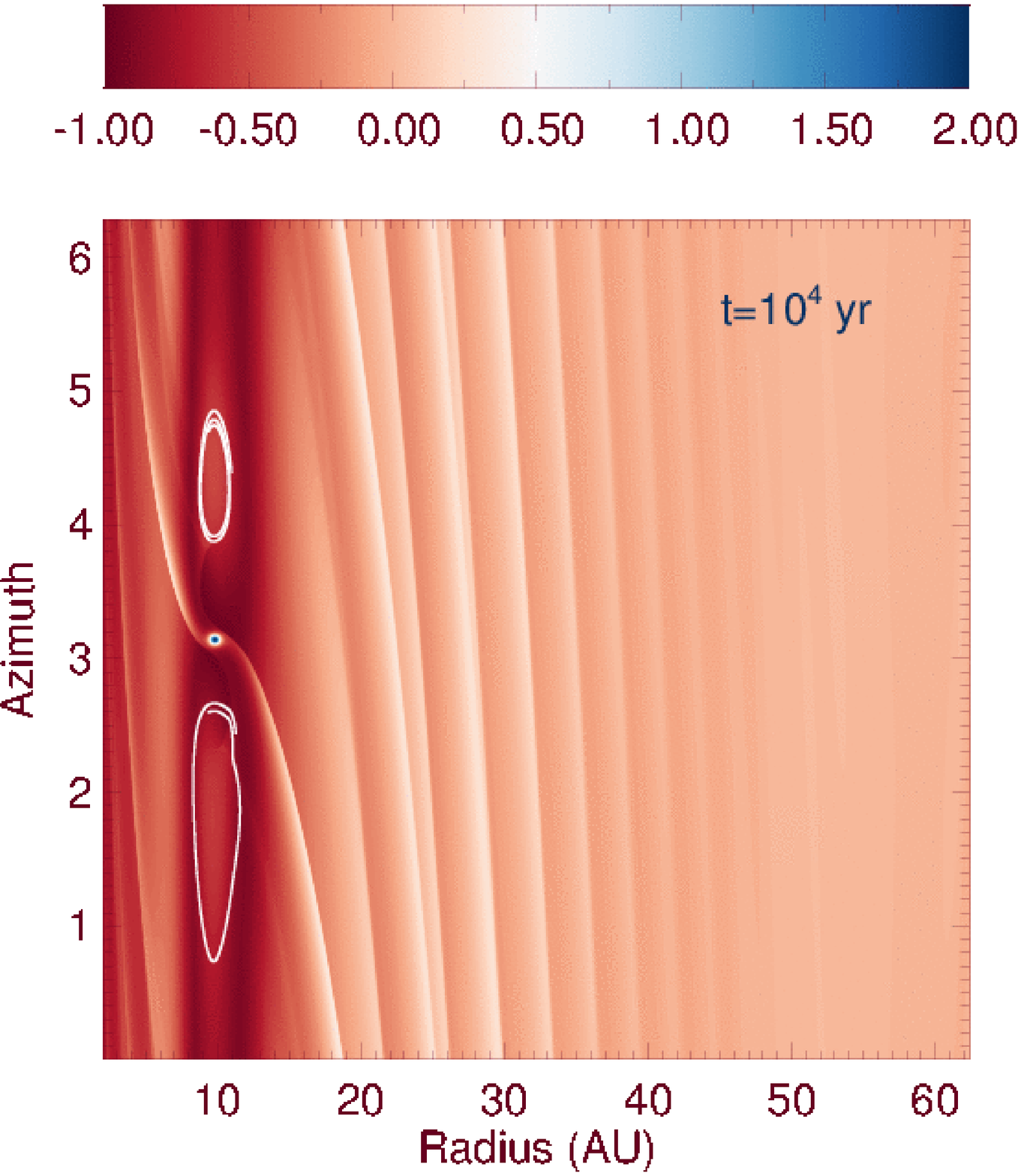}

\caption{Contours of the perturbed surface density at different times for the simulation with ${\dot M}=7\times 10^{-8}\;M_\odot/yr$ and 
in which a protoplanet with initial mass of $20M_\oplus$ accretes gas material on characteristic timescale 
$t_{acc}=5t_{dyn}$. }
\label{fig:2dmaps}
\end{figure*}

\subsection{Conditions for triggering fast outward migration}
As mentioned above, $20-30$ $M_\oplus$ are expected to undergo runaway gas accretion in the standard core accretion scenario (Pollack et al. 1996).  
In this section, we examine how gas accretion affects the orbital evolution of a planet that is 
subject to strong dynamical corotation torques. Fig. \ref{fig:accret} presents  the results of simulations of the dynamical evolution of an 
accreting protoplanet with initial mass of $20$ $M_\oplus$, and that employ the  routine for gas accretion 
described in Sect. \ref{sec:initial}. 

Fig. \ref{fig:accret} shows that gas accretion strengthens dynamical torques at early times.  We indeed observe that an 
accreting protoplanet can experience large excursions in the outer disc, depending on the accretion rate ${\dot M}$ and the accretion parameter 
$f_{red}$. For example, in the run with ${\dot M}=10^{-7}\;M_\odot/yr$ and $f_{red}=10$, an accreting core starting at $a=11$ AU  
rapidly migrates outward to $a\sim 75  $ AU under the combined effects of dynamical torques and gas accretion. In presence of 
gas accretion, there is also evidence for  runaway outward migration in accretion discs models with ${\dot M}=5\times 10^{-8}\;M_\odot/yr$  whereas runs in which gas accretion was neglected showed fast outward migration for ${\dot M} \ge 7\times 10^{-8}\;M_\odot/yr$ only.  \\

It is plausible that the positive effect of gas accretion on the strength of dynamical torques is  related to the onset of non-linearities as the planet 
mass ratio $q$ approaches $q\sim h_p^3$. Results from previous work 
(Paardekooper 2014; Pierens 2015) indeed indicate that dynamical torques can easily make intermediate mass planets with $q\sim h_p^3$ enter the 
outward runaway migration regime. This occurs because such planets tend to have a larger horseshoe region than low-mass cores (Masset et al. 2006), 
resulting in a boost of the corotation torque associated with  the gas material that flows across the planet orbit. \\
As discussed in Pierens (2015), fast outward migration is triggered whenever a planet's drift rate ${\dot a}$ is 
faster than the critical drift rate ${\dot a}_f=3x_s^2\Omega_p/8\pi a$, such that the planet migrates a distance greater than the horseshoe 
width over one horseshoe libration time. The migration rate ${\dot a}$  of the planet is given by:

\begin{equation}
{\dot a}=\frac{2}{\pi} (\Gamma/\Gamma_0) \frac{qq_d}{h_p^2} a \Omega_p
\end{equation}

where $q_d=\pi \Sigma_p a^2$ corresponds to  the mass of gas material contained inside the planet orbit. For low-mass planets,  the criterion for runaway migration given by Eq. \ref{eq:calq} can be 
recovered by simply substituting an estimation for the half-width of the horseshoe region  $x_s= 1.2 a \sqrt{q/h_p}$ in the condition ${\dot a} > {\dot a}_f$. For higher mass planets with 
$x_s\sim 2.45 a (q/3)^{1/3}$ (Masset et al. 2006), however, it can be shown  that the condition 
 ${\dot a} > {\dot a}_f$ is equivalent to:

\begin{equation}
{\cal Q} \lesssim \frac{2}{\gamma}\left(\frac{\gamma \Gamma}{\Gamma_0}\right)q^{1/3}h_p^{-1}
\label{eq:calq1}
\end{equation}

Interestingly, we see that the condition for the planet to enter the fast migration regime now depends on the planet mass ratio $q$ and that it is 
easier to get runaway migration as the planet grows. In principle, the term $\Gamma/\Gamma_0$ also depends on the planet 
mass ratio since the disc surface density profile is expected to be altered if the planet mass is high enough. In a recent work, 
however, Lega et al. (2015) found that even for planets that are subject to significant non-linear effects, there is still quite a good agreement between the torque obtained from 3D hydrodynamical  simulations and the one 
predicted by existing analytical formula. This  indicates that in accreting disc models,  the term  $\Gamma/\Gamma_0$ is almost independent of $q$ for intermediate planet masses with $q\sim h_p^3$.  We also note that the condition given by Eq. \ref{eq:calq1} implicitly depends on the softening 
parameter that is adopted for the planet potential. Since the entropy-related horseshoe drag scales as $(b/H_p)^{-1}$ (e.g. Baruteau 
\& Masset 2013), more massive 
discs (or equivalently higher accretion rates)   would be required to  trigger runaway migration for values of $b/H_p$ larger than the one 
considered here.\\
For a given value of the Toomre parameter ${\cal Q}$, Eq. \ref{eq:calq1} provides an estimation of  the minimum value for the mass ratio $q_{min}$ above which fast 
migration is expected to occur. It is given by:
\begin{equation}
q_{min}=\frac{\gamma^3}{8}\left(\frac{\gamma \Gamma}{\Gamma_0}\right)^{-3}({\cal Q} h_p)^3
\label{eq:qmin}
\end{equation}
Considering the model with ${\dot M}=7\times10^{-8}\;M_\odot/yr$, the middle panel of Fig. \ref{fig:accret} indicates that runaway migration is triggered once the planet 
is located at $a\sim 10$ AU and its mass ratio has reached $q\sim 2\times 10^{-4}$. Assuming 
that the corotation torque is unsaturated we have  (Paardekooper et al. 2010):
\begin{equation}
\gamma \Gamma/\Gamma_0=-2.5-1.7\beta+0.1\sigma+7.9\frac{\xi}{\gamma}
\end{equation}
where $\beta$ and $\xi$ are the negative of the power-law indices associated with the temperature and entropy profiles respectively. For 
${\dot M}=7\times10^{-8}$, we obtain $\sigma=0.16$, $\beta=1.35$, $\xi=1.29$ at $R\sim 10$ AU and consequently $\gamma \Gamma /\Gamma_0 \sim 2.5$. Moreover, examination of Fig. \ref{fig:struct} shows that ${\cal Q}\sim 3$ and $h\sim 0.06$ at $R\sim 10$ AU. Inserting  these values 
into Eq. \ref{eq:qmin} leads to 
$q_{min}\sim 1.3\times 10^{-4}$, which is in reasonable agreement with the results of the calculation corresponding to ${\dot M}=7\times10^{-8}\;M_\odot/yr$. \\

Fig. \ref{fig:accret} shows that the ability of an accreting core to undergo runaway outward 
migration  is a sensitive 
function of the accretion parameter $f_{red}$. In the case where the growth rate is maximum, namely for $f_{red}=1$,  the 
resulting strong Lindblad torque rapidly leads to a reversal of the migration direction. In the opposite limit of slow gas accretion, however, 
corotation torque saturation can prevent the planet from entering the fast migration regime.  A crude estimation of the minimum value 
for $f_{red}$ that is required to avoid saturation of the corotation torque as the planet accretes gas material from the disc can be 
obtained from the argument that the doubling-mass timescale $t_d$ needs to be shorter than the libration timescale $t_{lib}$. For our
accretion procedure (see Sect . \ref{sec:initial}), the doubling mass timescale is given by:
\begin{equation}
t_d\sim 6\frac{q^{1/3}}{q_d}t_{acc}
\end{equation}
Substituting this previous expression in the criterion $t_d<t_{lib}$ and using $x_s\sim a\sqrt{q/h_p} $, we find that a necessary condition 
for the planet to avoid the effect of corotation torque saturation as it grows is given by:
\begin{equation}
f_{red}\lesssim 0.2q_dq^{-1/3}h_p^{1/2}
\label{eq:fred}
\end{equation}
If we consider our nominal disc model with ${\dot M}=7\times 10^{-8}\;M_\odot/yr$, $q_d\sim 0.018$ at $R=10$ AU, which leads to 
the prediction that a protoplanet with initial mass ratio $q=6\times 10^{-5}$ will undergo strong dynamical torques provided 
that $f_{red}\lesssim 3$. This is slightly smaller than what we report from the simulations, which resulted in fast outward migration  in the runs with $f_{red}\le 5$, as can be observed in Fig. \ref{fig:accret}.

\subsection{Long term evolution}

\begin{figure}
\centering
\includegraphics[width=\columnwidth]{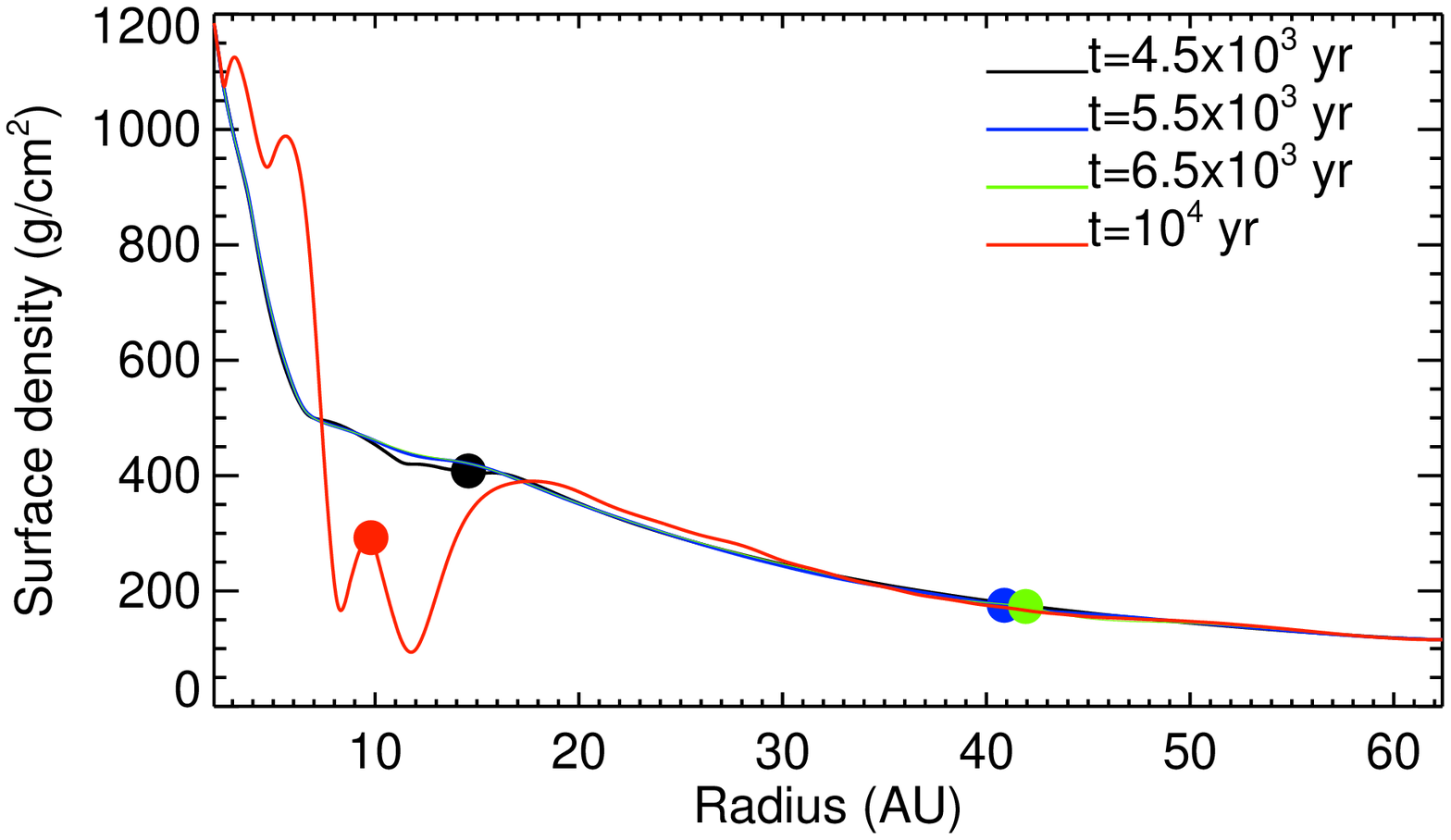}
\includegraphics[width=\columnwidth]{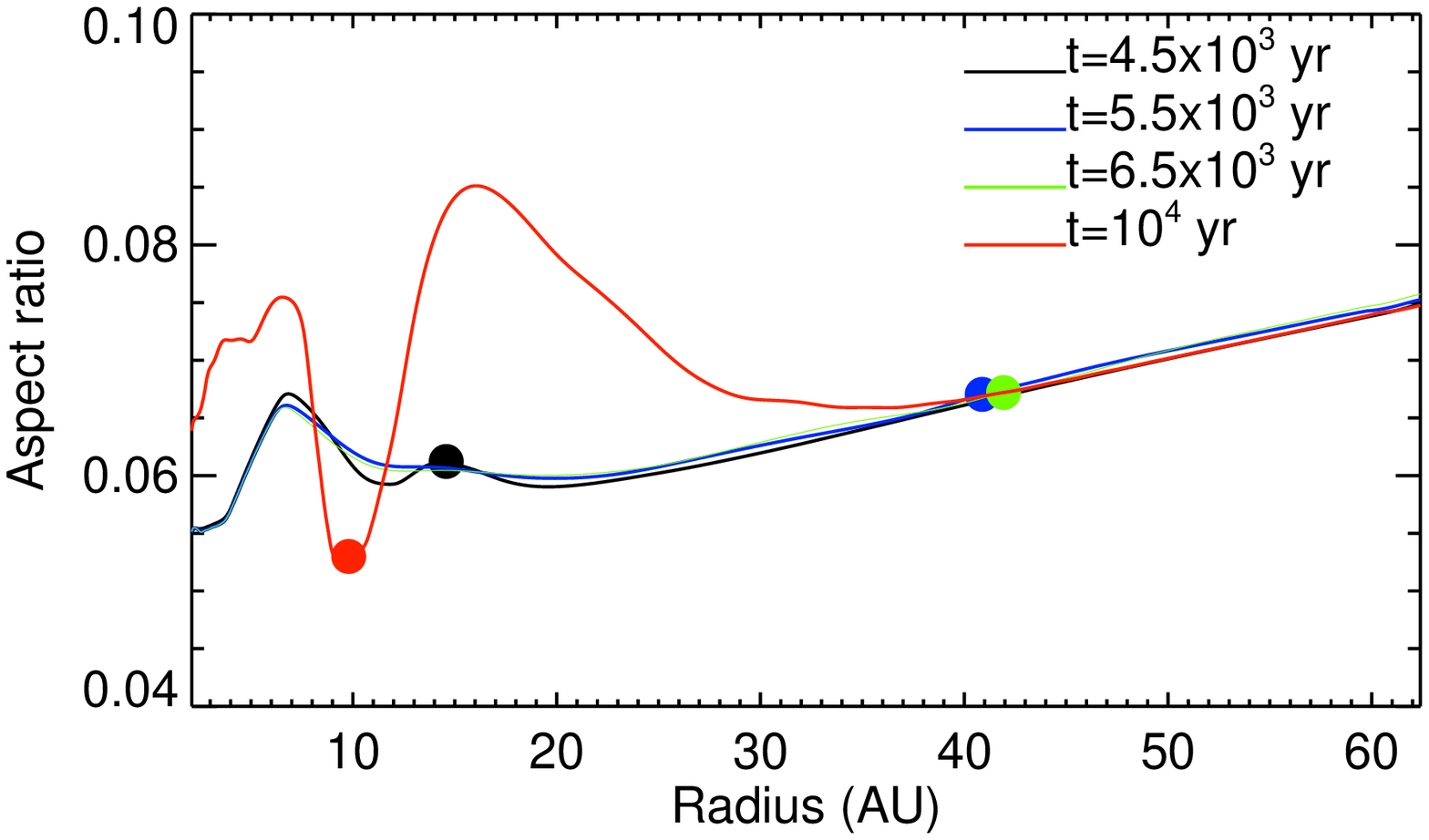}
\caption{Disc surface density (upper panel) and aspect ratio (lower panel) profiles at different times for the run  with ${\dot M}=7\times 10^{-8}\;M_\odot/yr$ and 
in which a protoplanet with initial mass of $20M_\oplus$ accretes gas material on characteristic timescale 
$t_{acc}=5t_{dyn}$.}
\label{fig:profiles}
\end{figure}

Although gas accretion can, under certain circumstances,  boost the effect of dynamical torques, it appears from 
Fig. \ref{fig:accret} that an episode of fast outward migration 
cannot be sustained over long timescales. This arises because of the following effects:\\
i) The negative Lindblad torque increases in strength with the planet mass. Because the mass contained 
within a planet's feeding zone scales as $m_{feed}\propto a R_H\Sigma(a)$ with $\Sigma(a)\propto a^{-15/14}$ (see Sect. \ref{sec:initial}) in the outer parts of the disc, outward migration promotes rapid growth of the planet, such that the Lindblad torque  
comes to dominate dynamical torques.  We note that for the highest accretion rates that are considered,  the mass growth 
at large planet masses should be in reality reduced due to a high accretion luminosity (Stamatellos 2015). \\
ii) The  trapped underdense region cools due to thermal diffusion, resulting in a slight increase in surface density.  The 
implication is that the negative feedback effect from that region on migration is enhanced as the planet migrates outward. \\
iii) Part of the surface density deficit within the trapped region is progressively  
lost as outward migration proceeds. This is illustrated in Fig. \ref{fig:2dmaps} where we display contour plots of the perturbed surface density at different times for the run with ${\dot M}=7\times 10^{-8}\;M_\odot/yr$ and 
$f_{red}=5$,.  At early 
times, the streamlines that are overplotted show gas flowing across the planet's orbit, together with the underdense librating material that is 
responsible for the strong dynamical torques experienced by the planet. At $t=5.5\times 10^3$ yr, however, we see that the gas material that is bound to the planet   
now librates around the $L5$ Lagrange point, and this tends to remove the surface density deficit in the underdense region. Interestingly, it has been shown that a similar effect is responsible for  stalling 
of runaway migration in the context of the classical Type III migration regime (Masset \& Papaloizou 2003; 
Peplinski et al. 2008; Lin \& Papaloizou 2010).  \\
The third panel of Fig. \ref{fig:2dmaps} shows the perturbed surface density at $t=6.5\times 10^3$ yr, just after the direction of migration has been reversed.  At that time, the planet mass ratio is $q\sim 7\times 10^{-4}$ (see Fig. \ref{fig:accret}) and its 
semi major axis $a\sim 40$ AU.  Because the planet  evolves  in a region of the disc whose thermodynamical state is close to the isothermal limit, the trapped region which is now located at the leading side of the planet has almost the same surface density as the ambient disc. Moreover, we see that the 
planet has not opened a gap at that time. The gap opening criterion of Crida et al. (2006) is given by:
\begin{equation}
1.1\left(\frac{h_p^3}{q}\right)^{1/3}+50\left(\frac{\alpha}{h_p}\right)\left(\frac{h_p^3}{q}\right)\lesssim 1
\label{eq:gap}
\end{equation}
which is not satisfied for $q=7\times 10^{-4}$ due to the high value for the aspect ratio $h_p\sim 0.065$ at the planet's location.  Gap clearing 
is therefore not expected for $q=7\times 10^{-4}$, consistent with what we observe in the simulation. This is further illustrated in  
Fig. \ref{fig:profiles}, which shows the disc surface density and aspect ratio profiles at different times for the same run . \\
The planet subsequently migrates back inward because of the strong tidal torque resulting from the large planet and inner disc masses. At 
$t\sim 10^4$ yr (the fourth panel of Fig. \ref{fig:2dmaps}) the planet has opened a gap in the disc, 
which is confirmed by looking at the corresponding disc surface density and aspect ratio plots in Fig. \ref{fig:profiles}.  We note that 
these profiles should be taken with care since self-shadowing effects, which are not included in this study (see Sect. \ref{sec:num} ), are expected to be important 
once the planet opens a deep gap in the disc.  From this time onward, 
Fig. \ref{fig:accret} shows that the planet  slowly migrates inward,  with a drift rate corresponding to the Type II regime.  We note that 
such a mode 
of evolution is  similar to what is seen in  the other runs where the planet experienced a large excursion in the outer disc. \\
 Compared to the prediction from the criterion given by Eq. \ref{eq:gap}, a common feature of these runs is that we find 
that gap opening occurs at slightly larger masses. For the reference run with ${\dot M}=7\times 10^{-8}\;M_\odot/yr$ and $f_{red}=5$, for example, the planet is actually found 
to open a gap once its mass ratio has reached  $q\sim 1.6\times 10^{-3}$ whereas Eq. \ref{eq:gap} predicts that this should occur for 
$q\sim 1.1\times 10^{-3}$. As demonstrated  by Malik et al. (2015), such a discrepancy is due to the fast inward migration of the planet and  arises because the planet migrates across the region 
where it is expected to open a gap on a timescale shorter than the gap opening timescale. \\ 

\section{Discussion and conclusion}

In this paper, we have presented the results of hydrodynamical simulations of accreting protoplanets embedded in radiative 
discs. The goal of this study was to examine the influence of gas accretion onto the orbital evolution of outward 
migrating planets subject to strong dynamical corotation torques. In relatively massive protoplanetary 
discs, dynamical torques can make  a low/intermediate mass planet  enter a fast migration regime and experience large excursions in the outer disc.  \\
   We considered accreting disc models with constant mass flow through the disc and where radiative cooling is balanced by 
viscous and stellar heating, and studied how dynamical torques depend on the accretion rate ${\dot M}$. For an 
$\alpha$ viscous stress parameter $\alpha=2\times 10^{-3}$, we find evidence of strong dynamical torques 
for $\dot{M}\ge 7\times10^{-8}\;M_\odot/yr$. In that case,   cores with fixed mass of $\sim$ 20-30 Earth masses  migrate outward with drift rates that are significantly larger than those 
predicted from analytical formulae for the disc torque, and are subsequently trapped at or just beyond the predicted zero-torque radius where the 
Lindblad and corotation torques cancel each other. Protoplanets with initial mass of  $20$ $M_\oplus$ and undergoing runaway gas accretion, however, are 
found  to enter a runaway outward migration regime under some circumstances. Depending on the values for the mass flux through the 
disc and the accretion timescale onto the planet,  protoplanets formed in the 5-10 AU region may migrate well beyond the the zero-torque radius out to separations of $\gtrsim 50\; AU$.  \\
We find that gas accretion has two opposite effects. It is easier for an accreting protoplanet to enter a fast migration regime due to the 
onset of non-linearities as its mass ratio approaches $q\sim h_p^3$, where $h_p$ is the disc aspect ratio at the planet location. Planet growth, however,  is accompanied by an  increase of the Lindblad torque which 
tends to counteract the effect of dynamical torques for an outward migrating planet. This, combined with the fact that the structure of the region that 
is bound to the planet  is significantly altered as a result of mass accretion and  diffusion processes, systematically  leads to a change in the direction 
of migration. Interestingly, it is worthwhile to notice that similar results have been reported from simulations of accreting planets undergoing outward-directed Type III 
migration in isothermal discs (Peplinski et al. 2008). \\
Reversal of migration typically occurs once the planet mass has reached $\sim 1M_J$.  Due to the high value for the disc aspect ratio at 
the planet location, a Jupiter mass 
planet that stops its outward migration does not open a gap in the outer parts of our accreting disc models.  This resuts in a 
very fast inward migration due to  the large planet and inner disc masses, which further prevents the planet from opening a gap.  Such a 
process can however arise at later times,  once the planet is massive enough for  the gap opening timescale to be shorter than the migration 
timescale. This corresponds to the transition to Type II migration, which typically occurs in our simulations at planet masses of $\sim 2M_J$ and orbital distances of $\sim 10$ AU. \\
 At first glance, its therefore seems difficult for such a mechanism  to  explain the presence of massive planets on wide orbits, and the population of giant planet residing beyond a few AU as well.  The work of Coleman \& Nelson (2014) indeed suggests that for a Jovian planet to survive at a few AU, it must  have initiated runaway gas accretion and Type II migration beyond $\sim15$ AU typically, and  late in the disc lifetime (see also Bitsch et al 2015b). This is because giant planet cores that accrete gas tend to undergo rapid inward migration due to corotation torque saturation before they open a gap.  It is clear that the mechanism presented here could potentially allow a growing core that  undergoes runaway accretion to reach  large distances in the outer disc and therefore avoid corotation torque saturation. Our results indicate that this could lead to the formation of Jupiter-mass planets at orbiting at $\sim 10$ AU.  However, since this occurs for typical accretion rates higher than  $\sim 7\times 10^{-8} M_\odot/yr$ only, it would seem difficult  for such planets to survive type II migration before the disc is dispersed through photo evaporation and viscous evolution. \\
 It should be noted, however, that the simulations presented here contain a number of simplifications.  Although 
 dynamical corotation torques play an important role on the planet orbital evolution for typical  values of the 
 Toomre stability parameter of ${\cal Q}\lesssim 3-4$, we did not consider 
 the effect of disc self-gravity. Self-gravity is expected to cause a shift of Lindblad resonances (Pierens \& Hur\'e 2005; Baruteau \& Masset 2008), 
 resulting in a slightly stronger Lindbad torque in comparison with the case where 
 self-gravity is not considered. For highest accretion rates  we considered, ${\cal Q}$ can even fall below the gravitational stability limit 
 in the very outer parts of the disc, possibly resulting in gravitoturbulence if the gas cooling timescale is not too short (Baruteau et al. 2011). Moreover, 
 we have only considered the single planet case. In the multiple planet case, however, planet-planet interactions may prevent the fast inward migration 
 episode that is observed in the simulations. We will  examine this issue in a future paper. 
 
\section*{Acknowledgments}
Computer time for this study was provided by the computing facilities MCIA (M\'esocentre de Calcul Intensif Aquitain) of the Universite de Bordeaux and by HPC resources of Cines under the allocation  c2016046957 made by GENCI (Grand Equipement National de Calcul Intensif).  We thank the Agence Nationale pour la Recherche under grant ANR-13-BS05-0003 (MOJO).


\begin{thebibliography}{}

\bibitem[Baruteau 
\& Masset(2008)]{2008ApJ...672.1054B} Baruteau, C., \& Masset, F.\ 2008, ApJ, 672, 1054
\bibitem[Baruteau et al.(2011)]{2011MNRAS.416.1971B} Baruteau, C., Meru, F., \& Paardekooper, S.-J.\ 2011, MNRAS, 416, 1971
\bibitem[Baruteau \& Masset(2013)]{2013LNP...861..201B} Baruteau, C., \& Masset, F.\ 2013, Lecture Notes in Physics, Berlin Springer Verlag, 861, 201
\bibitem[Bell 
\& Lin(1994)]{1994ApJ...427..987B} Bell, K.~R., \& Lin, D.~N.~C.\ 1994, ApJ, 427, 987
\bibitem[Bitsch et 
al.(2014)]{2014A&A...564A.135B} Bitsch, B., Morbidelli, A., Lega, E., \& Crida, A.\ 2014, A\&A, 564, A135
\bibitem[Bitsch et al.(2015)]{2015A&A...575A..28B} Bitsch, B., Johansen, A., Lambrechts, M., \& Morbidelli, A.\ 2015, A\&A, 575, A28
\bibitem[Bitsch et al.(2015)]{2015A&A...582A.112B} Bitsch, B., Lambrechts, M., \& Johansen, A.\ 2015b, A\& A, 582, A112
\bibitem[Brandt et al.(2014)]{2014ApJ...794..159B} Brandt, T.~D., McElwain, M.~W., Turner, E.~L., et al.\ 2014, ApJ, 794, 159
\bibitem[Chiang \& Goldreich(1997)]{1997ApJ...490..368C} Chiang, E.~I., \& Goldreich, P.\ 1997, ApJ, 490, 368
\bibitem[Coleman \& Nelson(2014)]{2014MNRAS.445..479C} Coleman, G.~A.~L., \& Nelson, R.~P.\ 2014, MNRAS, 445, 479 
\bibitem[Crida et al.(2006)]{2006Icar..181..587C} Crida, A., Morbidelli, A., \& Masset, F.\ 2006, Icarus, 181, 587
\bibitem[Crida et al.(2009)]{2009ApJ...705L.148C} Crida, A., Masset, F., \& Morbidelli, A.\ 2009a, ApJL, 705, L148
\bibitem[Crida et al.(2009)]{2009A&A...502..679C} Crida, A., Baruteau, C., Kley, W., \& Masset, F.\ 2009b, A\&A, 502, 679
\bibitem[Cumming et al.(2008)]{2008PASP..120..531C} Cumming, A., Butler, 
R.~P., Marcy, G.~W., et al.\ 2008, PASP, 120, 531
\bibitem[D'Angelo 
\& Lubow(2008)]{2008ApJ...685..560D} D'Angelo, G., \& Lubow, S.~H.\ 2008, ApJ, 685, 560
\bibitem[de Val-Borro et al.(2006)]{2006MNRAS.370..529D} de Val-Borro, M., Edgar, R.~G., Artymowicz, P., et al.\ 2006, MNRAS, 370, 529
\bibitem[Dodson-Robinson et al.(2009)]{2009ApJ...707...79D} Dodson-Robinson, S.~E., Veras, D., Ford, E.~B., \& Beichman, C.~A.\ 2009, ApJ, 707, 79
\bibitem[Fung et al.(2015)]{2015ApJ...811..101F} Fung, J., Artymowicz, P., \& Wu, Y.\ 2015, ApJ, 811, 101 
\bibitem[Go{\'z}dziewski \& Migaszewski(2014)]{2014MNRAS.440.3140G} Go{\'z}dziewski, K., \& Migaszewski, C.\ 2014, MNRAS, 440, 3140
\bibitem[Hayashi(1981)]{1981PThPS..70...35H} Hayashi, C.\ 1981, Progress of 
Theoretical Physics Supplement, 70, 35
\bibitem[Kley(1989)]{1989A&A...208...98K} Kley, W.\ 1989, A\&A, 208, 98
\bibitem[Kley(1999)]{1999MNRAS.303..696K} Kley, W.\ 1999, MNRAS, 303, 696
\bibitem[Kley 
\& Crida(2008)]{2008A&A...487L...9K} Kley, W., \& Crida, A.\ 2008, A\&A, 487, L9
\bibitem[Kley \& Haghighipour(2014)]{2014A&A...564A..72K} Kley, W., \& Haghighipour, N.\ 2014, A\&A, 564, A72
\bibitem[Lambrechts et al.(2014)]{2014A&A...572A..35L} Lambrechts, M., Johansen, A., \& Morbidelli, A.\ 2014, A\&A, 572, A35

\bibitem[Lega et al.(2015)]{2015MNRAS.452.1717L} Lega, E., Morbidelli, A., Bitsch, B., Crida, A., \& Szul{\'a}gyi, J.\ 2015, MNRAS, 452, 1717
\bibitem[Lin \& Papaloizou(2010)]{2010MNRAS.405.1473L} Lin, M.-K., \& Papaloizou, J.~C.~B.\ 2010, MNRAS, 405, 1473
\bibitem[Maire et al.(2015)]{2015A&A...576A.133M} Maire, A.-L., Skemer, A.~J., Hinz, P.~M., et al.\ 2015, A\&A, 576, A133 
\bibitem[Malik et al.(2015)]{2015ApJ...802...56M} Malik, M., Meru, F., Mayer, L., \& Meyer, M.\ 2015, ApJ, 802, 56
\bibitem[Masset(2000)]{2000A&AS..141..165M} Masset, F.\ 2000, A\&AS, 141, 165
\bibitem[Masset 
\& Papaloizou(2003)]{2003ApJ...588..494M} Masset, F.~S., \& Papaloizou, J.~C.~B.\ 2003, ApJ, 588, 494
\bibitem[Masset et al.(2006)]{2006ApJ...652..730M} Masset, F.~S., D'Angelo, 
G., \& Kley, W.\ 2006, ApJ, 652, 730
\bibitem[Marois et al.(2008)]{2008Sci...322.1348M} Marois, C., Macintosh, B., Barman, T., et al.\ 2008, Science, 322, 1348
\bibitem[Marois et al.(2010)]{2010Natur.468.1080M} Marois, C., Zuckerman, B., Konopacky, Q.~M., Macintosh, B., \& Barman, T.\ 2010, Nature, 468, 1080
\bibitem[Mayor et al.(2011)]{2011arXiv1109.2497M} Mayor, M., Marmier, M., Lovis, C., et al.\ 2011, arXiv:1109.2497
\bibitem[Menou 
\& Goodman(2004)]{2004ApJ...606..520M} Menou, K., \& Goodman, J.\ 2004, ApJ, 606, 520
\bibitem[Paardekooper et al.(2010)]{2010MNRAS.401.1950P} Paardekooper, S.-J., Baruteau, C., Crida, A., \& Kley, W.\ 2010, MNRAS, 401, 1950 
\bibitem[Paardekooper et al.(2011)]{2011MNRAS.410..293P} Paardekooper, 
S.-J., Baruteau, C., \& Kley, W.\ 2011, MNRAS, 410, 293
\bibitem[Paardekooper(2014)]{2014MNRAS.444.2031P} Paardekooper, S.-J.\ 
2014, MNRAS, 444, 2031
\bibitem[Pepli{\'n}ski et al.(2008)]{2008MNRAS.386..179P} Pepli{\'n}ski, 
A., Artymowicz, P., \& Mellema, G.\ 2008, MNRAS, 386, 179
\bibitem[Pierens 
\& Hur{\'e}(2005)]{2005A&A...433L..37P} Pierens, A., \& Hur{\'e}, J.-M.\ 2005, A\&A, 433, L37
\bibitem[Pierens(2015)]{2015MNRAS.454.2003P} Pierens, A.\ 2015, MNRAS, 454, 2003
\bibitem[Shvartzvald et al.(2016)]{2016MNRAS.457.4089S} Shvartzvald, Y., Maoz, D., Udalski, A., et al.\ 2016, MNRAS, 457, 4089
\bibitem[Stamatellos(2015)]{2015ApJ...810L..11S} Stamatellos, D.\ 2015, ApJL, 810, L11
\bibitem[Pollack et al.(1996)]{1996Icar..124...62P} Pollack, J.~B., Hubickyj, O., Bodenheimer, P., et al.\ 1996, Icarus, 124, 62
\bibitem[van Leer(1977)]{1977JCoPh..23..276V} van Leer, B.\ 1977, Journal 
of Computational Physics, 23, 276 

\end{thebibliography}
\end{document}